\documentclass[twocolumn,superscriptaddress,prd,floatfix]{revtex4-1} 

\pdfoutput=1 

\usepackage{verbatim}
\usepackage{mathtools}
\usepackage{relsize}
\usepackage{natbib}

\usepackage{amsmath}
\usepackage{graphicx}
\usepackage[capposition=bottom]{floatrow}
\usepackage{amssymb}
\usepackage{rotating}
\usepackage{afterpage}
\usepackage{float}
\usepackage{geometry}
\usepackage{titlesec}
\usepackage{epstopdf}

\usepackage{latexsym}
\usepackage{color}
\usepackage{array}
\usepackage{wasysym}
\usepackage{verbatim}

\renewcommand\thesection{\Roman{section}.}
\renewcommand\thesubsection{\arabic{subsection}}
\renewcommand\thesubsubsection{\alph{subsubsection}}
\titleformat{\section}[block]{\bfseries\filcenter}{\thesection}{1em}{}
\titleformat{\subsection}[block]{\bfseries\filcenter}{\thesubsection.}{1em}{}
\titleformat{\subsubsection}{\bfseries\filcenter}{\thesubsubsection.}{1em}{}

\usepackage{titlesec}

\begin{document}

\title{Improved Measurements of the $\beta$-Decay Response of Liquid Xenon with the LUX Detector}

\author{D.S.~Akerib} \affiliation{Case Western Reserve University, Department of Physics, 10900 Euclid Ave, Cleveland, OH 44106, USA} \affiliation{SLAC National Accelerator Laboratory, 2575 Sand Hill Road, Menlo Park, CA 94205, USA} \affiliation{Kavli Institute for Particle Astrophysics and Cosmology, Stanford University, 452 Lomita Mall, Stanford, CA 94309, USA}
\author{S.~Alsum} \affiliation{University of Wisconsin-Madison, Department of Physics, 1150 University Ave., Madison, WI 53706, USA}  
\author{H.M.~Ara\'{u}jo} \affiliation{Imperial College London, High Energy Physics, Blackett Laboratory, London SW7 2BZ, United Kingdom}  
\author{X.~Bai} \affiliation{South Dakota School of Mines and Technology, 501 East St Joseph St., Rapid City, SD 57701, USA}  
%\author{A.J.~Bailey} \affiliation{Imperial College London, High Energy Physics, Blackett Laboratory, London SW7 2BZ, United Kingdom}  
\author{J.~Balajthy} \affiliation{University of Maryland, Department of Physics, College Park, MD 20742, USA}\affiliation{University of California Davis, Department of Physics, One Shields Ave., Davis, CA 95616, USA}  
\author{A.~Baxter} \affiliation{University of Liverpool, Department of Physics, Liverpool L69 7ZE, UK}  
\author{P.~Beltrame} \affiliation{SUPA, School of Physics and Astronomy, University of Edinburgh, Edinburgh EH9 3FD, United Kingdom}  
\author{E.P.~Bernard} \affiliation{University of California Berkeley, Department of Physics, Berkeley, CA 94720, USA}  
\author{A.~Bernstein} \affiliation{Lawrence Livermore National Laboratory, 7000 East Ave., Livermore, CA 94551, USA}  
\author{T.P.~Biesiadzinski} \affiliation{Case Western Reserve University, Department of Physics, 10900 Euclid Ave, Cleveland, OH 44106, USA} \affiliation{SLAC National Accelerator Laboratory, 2575 Sand Hill Road, Menlo Park, CA 94205, USA} \affiliation{Kavli Institute for Particle Astrophysics and Cosmology, Stanford University, 452 Lomita Mall, Stanford, CA 94309, USA}
% ...

\begin{abstract}
We report results from an extensive set of measurements of the $\beta$-decay response in liquid xenon. These measurements are derived from high-statistics calibration data from injected sources of both $^{3}$H and $^{14}$C in the LUX detector. The mean light-to-charge ratio is reported for 13 electric field values ranging from 43~to~491~V/cm, and for energies ranging from 1.5~to~145~keV. 
\end{abstract}

%\twocolumn[\maketitle]
\maketitle

\section{Introduction}\label{section:Intro}
The Large Underground Xenon experiment (LUX) was a liquid-xenon (LXe) time-projection chamber (TPC). Before it was decommissioned in 2016, LUX was located in the Davis cavern of the Sanford Underground Research Facility (SURF) in Lead, South Dakota, on the 4,850' level~\cite{lux_2012}. In total, it contained about 370~kg of xenon, 250~kg of which was active. Energy deposits in the sensitive volume were detected using two arrays of 61 photomultiplier tubes (PMTs) at the top and bottom of the detector. 

LUX was initially designed as a WIMP dark-matter detector. The LUX full exposure of 3.35~$\times \ 10^4$~kg~days combines the first data-taking run (WS2013), which took place from April to August~2013~\cite{lux_2014,lux_2016}, with the second data-taking run (WS2014-2016), which ran from September~2014 until May~2016~\cite{lux_2017}. A 50~GeV/c$^2$ WIMP with a cross section greater than 1.1~$\times \ 10^{-46}$~cm$^2$ is excluded with 90\% confidence. Recently, stronger limits have been placed by the XENON-1T and PandaX experiments~\cite{xenon_1t,pandax}.

As a two-phase TPC, LUX was sensitive to light and charge signals via primary ($S1$) and secondary ($S2$) scintillation, respectively. The light and charge yields ($Ly$ and $Qy$) are defined as the average number of quanta (photons and electrons) per keV of energy deposited in the LXe. These depend upon the energy of the event, the magnitude of the electric field applied at the event's location, and whether the interaction leads to a nuclear recoil (NR) or an electron recoil (ER). The yields of an electron recoil may also depend upon further specifics of the interaction. For instance, interactions of $\beta$-particles in LXe may produce different yields from those of gammas, because the latter has some energy-dependent probability of photoabsorption, while the former does not~\cite{nest2}. Such variations will be seen in section~\ref{sec:discussion} when comparing values of $Qy$ from $\beta$ interactions with those from $^{83m}$Kr and $^{131m}$Xe decays.

The most prevalent and problematic backgrounds in current and future LXe dark matter experiments are $\beta$-decays of Rn daughters~\cite{lz_tdr,lz_sensitivity,xenon_1t,pandax}. It is therefore important to understand the $\beta$-decay-induced light and charge yields in LXe as a function of electric field and energy. Previous measurements of these yields using LUX WS2013 data, including a set of measurements using a $^{3}$H injection source at both 105~V/cm and 180~V/cm were reported in Ref.~\cite{lux_tritium,DQyields,Evanyields}. In this article we use the data from a novel $^{14}$C injection and a high statistics $^{3}$H calibration, which were conducted after WS2014-2016, to extend the previous results over a much wider range of energies and electric fields. 

The electric field in the WS2014-2016 LUX detector was highly non-uniform, ranging from less than 50~V/cm to over 500~V/cm. In this article we divide the detector into thirteen electric field bins, with central values spanning from 43~to~491~V/cm, and obtain measurements of the light and charge yields for each associated field value~\cite{thesis}. The use of a $^{14}$C injection source in addition to $^{3}$H increases the energy range of our measurements by nearly an order of magnitude. The radioactive isotope $^{14}$C $\beta$-decays to the ground state of $^{14}$N with a Q-value of 156~keV, which is 8.6 times greater than the $^{3}$H Q-value of 18.1 keV~\cite{C14_Kuzminov,C14_Wietfeldt,lux_tritium}. 

The $^{14}$C calibration was performed at the end of the LUX operational lifetime and just before decommissioning in September,~2016. We will refer to this period as ``post-WS''. The activities used in post-WS calibrations were allowed to be significantly greater than previous calibrations because maintaining low detector backgrounds was no longer a requirement. This resulted in a data set of roughly 2 million $^{14}$C events. After fiducial cuts, each of the thirteen electric field bins has between 60,000 and 120,000 events. A separate post-WS $^{3}$H dataset is also analyzed, which has about one third of the number of events as the $^{14}$C set. 

\section{Data Selection}
An interaction in the sensitive LXe produces primary scintillation photons and ionization electrons. The primary scintillation is collected by the PMTs and constitutes the $S1$ signal. The electrons are transported through an electric drift field to the top of the detector, where the electrons are extracted from the liquid surface into a region of gaseous xenon and produce the $S2$ signal through electroluminescence. The $S2$ signal is proportional to the number of electrons extracted.

The low-energy electronic depositions studied in this work have very short track lengths ($\sim$0.3~mm)~\cite{lxe_detectors}, and are treated as occurring at points in space. The signals caused by these depositions will therefore have a single $S1$ followed by a single $S2$. The $S2$ light generated by an extracted electron is highly localized in the x-y plane at the top of the detector, so the x-y position of an event can be reconstructed by analyzing the relative size of the pulses in the top PMT array. The drifting electrons take many microseconds longer to reach the liquid surface than the $S1$ photons take to be detected. The resulting difference in time between the $S1$ and $S2$ is referred to as the ``drift time''. In a detector with a uniform electric drift field, the electrons drift vertically at a constant velocity, so drift time gives a direct measurement of the z-position of an event~\cite{lux_posrec}.

The charge and light collection efficiencies have some dependence upon position due to the attachment of drifting electrons to impurities and detector geometry. These effects are measured using the response to $^{83m}$Kr and $^{3}$H~\cite{lux_2017}. A new set of efficiency-corrected data has been produced in which these effects are corrected for. In this article, $S1$ and $S2$ refer to these corrected values unless otherwise specified. 

To avoid edge effects in our analysis, we reject events near the boundaries of the sensitive volume. Events within about 3~cm from the walls of the detector were rejected using a radial cut which is described in section~4.2.2 of Ref.~\cite{thesis}. For the same reason, events with drift times greater than 330~$\mu$s or less than 10~$\mu$s were also rejected. 

The simplest selection cut used to isolate single site events is to reject any event that does not contain exactly one $S1$ and one $S2$, with the $S1$ occurring before the $S2$. The efficiency of this cut is found to decrease at higher energies due to correlated pile-up in both $S1$ and $S2$. In this work we use a modified version of this cut which is described in detail in section~4.2.1 of Ref.~\cite{thesis}. We require a selected event have at least one $S2$, and at least one $S1$ before the first $S2$. The first $S1$ and $S2$ pulses are required to contain at least 93\% of the total $S1$ and $S2$ area, respectively. This modified selection cut increases the acceptance of $^{131m}$Xe events from 61\% to 92\%, and improves the acceptance of $^{14}$C events to more than 90\% across the entire energy spectrum~\cite{thesis}. We use $^{131m}$Xe as a test of our selection cut because it is a mono-energetic peak at 163.9~keV, just above the $^{14}$C Q-value. The background rate remains very small in comparison to the injected sources, so the additional leakage of noise events due to the relaxed cut is negligible.

During and after WS2014-2016, the electric fields in LUX were highly non-uniform. A comprehensive study of the drift-field was performed and is detailed in Ref.~\cite{lux_efield}. This study produced high-resolution maps of the electric field. Using a 3-dimensional linear interpolation of these maps, we assign a specific field value to every event in our data sets. This enables us to define 13 bins in electric field whose centers range from 43~to~491~V/cm, and where each bin extends 10\% above and below its central value. We also limit the range of drift times that are drawn from so that a bin does not extend past the central drift time values of its adjacent bins. These bins are described in greater detail in section~4.2.3 of Ref.~\cite{thesis}.

\section{Calibration Source Injections}
\begin{figure}[b]
\centering
\includegraphics[width=\linewidth]{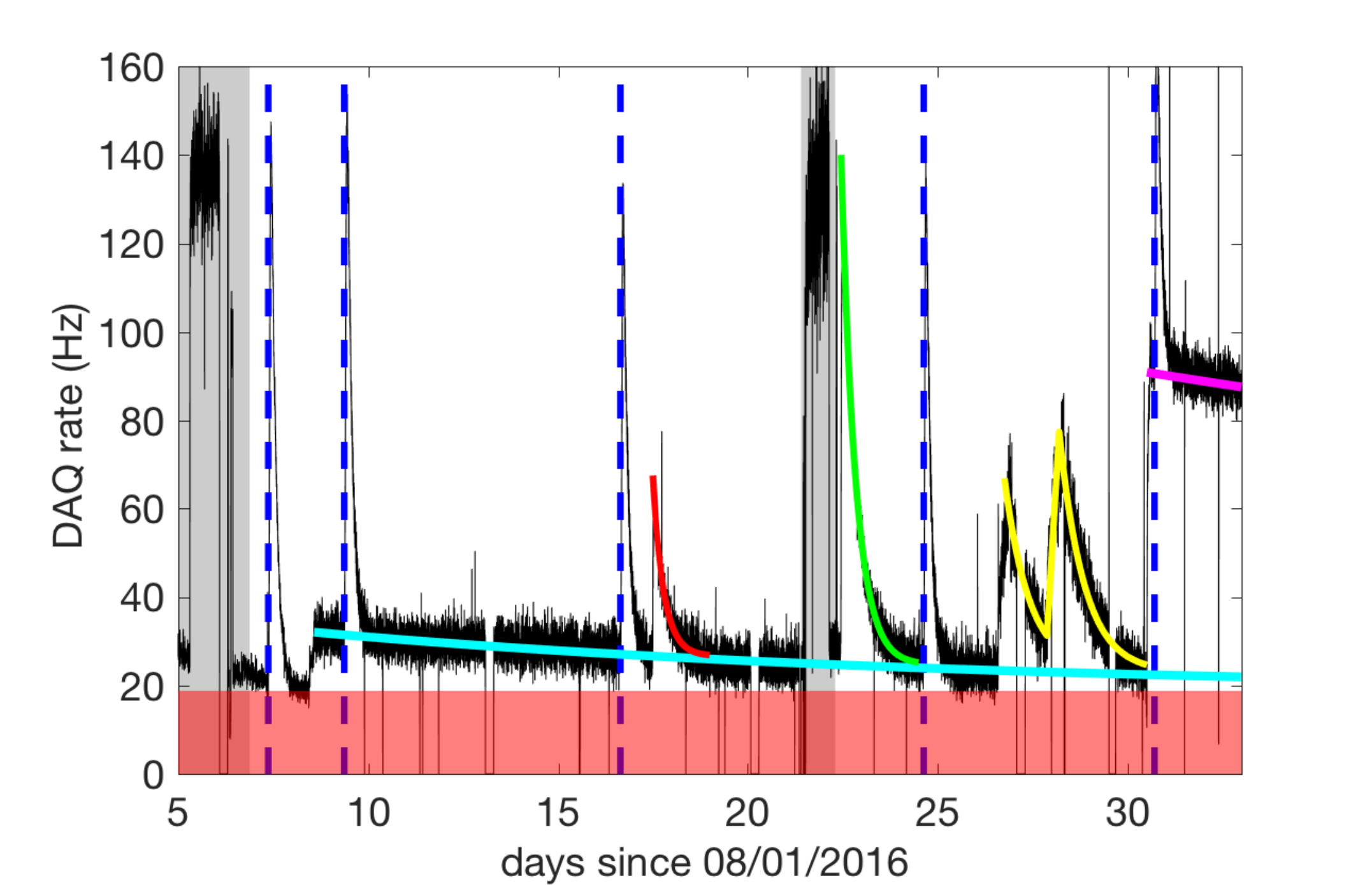}
\caption{The black line shows the data acquisition rate for the post-WS injection campaign. The red shaded region shows the constant baseline, while the grey shaded regions show the DD NR-calibration campaign. The vertical dashed blue lines show the times of standard $^{83m}$Kr injections. The cyan, red, green, and magenta lines trace the activities following the $^{131m}$Xe, $^{3}$H, $^{14}$C, and $^{37}$Ar, respectively. Additionally, the yellow lines indicate injections of $^{220}$Rn, which are not detailed in the text.} 
\captionsetup{justification=justified} 
\label{fig:DAQrate}
\end{figure}

After WS2014-2016 was completed in June of 2016, the detector was exercised with a variety of ER and NR calibration sources. The usual ER calibrations of $^{83m}$Kr~\cite{lux_kr1,lux_kr2,lux_kr3} and tritiated methane (CH$_3$T)~\cite{lux_tritium} were performed, along with NR calibrations using the deuterium-deuterium (DD) neutron generator~\cite{lux_dd1,lux_dd2}. In addition to these standard calibrations, novel techniques and sources were implemented. The timeline and activities of these injections can be seen in Figure~\ref{fig:DAQrate}.

\subsection{$^{131m}$Xe and $^{37}$Ar}
The isotope $^{131m}$Xe de-excites through a gamma transition to its ground state with an energy of 163.93~keV and a half life of 11.84~days. It is generated using a commercially available $^{131}$I pill and is introduced into the primary xenon circulation path using the $^{3}$H injection system described in Ref.~\cite{lux_tritium}. 

An injection of $^{37}$Ar was also deployed. This isotope decays via electron capture to $^{37}$Cl with a half-life of 35 days. In 90\% of these decays, a K-shell electron is captured, followed by the emission of x-rays and Auger electrons which total to 2.82~keV. There are also non-zero branching ratios for the capture of L- and M-shell electrons. $^{37}$Ar can be produced through stimulated $\alpha$ emission of a $^{40}$Ca target using a neutron beam. The $^{37}$Ar sample used in LUX was produced by irradiating an aqueous solution of CaCl$_2$ with neutrons from an AmBe source. The gas above this solution was then collected and purified to obtain the gaseous sample of $^{37}$Ar~\cite{pixey_ar37}. This sample was injected into the LUX xenon circulation using the same system as the $^{83m}$Kr calibrations~\cite{lux_kr2}.

\subsection{$^{14}$C and $^{3}$H}
The $^{3}$H injection system and procedure was described in detail in Ref.~\cite{lux_tritium}. In the post-WS injection campaign, it was used to deploy a high statistics $^{3}$H injection, as well as a novel $^{14}$C injection into the LUX detector. The $^{14}$C was in the form of radio-labeled methane which is chemically identical to the tritiated-methane used in Ref.~\cite{lux_tritium} and was also synthesized by Moravek Biochemical~\cite{moravek}. The methane, and therefore the $^{14}$C activity, is removed from the LUX xenon in the same manner as $^{3}$H via circulation through a heated zirconium getter. 

\section{Model of the LUX Post-WS $\beta$-Decay Data}
\subsection{Smearing of Continuous $\beta$ Spectra}\label{sec:desmear}
Measurements of energy-dependent parameters from continuous $\beta$ spectra are affected in nontrivial ways by finite detector resolution. We obtain measurements of yields and recombination by dividing the $^{14}$C and $^3$H into reconstructed energy bins. Finite detector resolution impacts our measurements by smearing some events into a reconstructed energy bin, whose true energies lie outside of the bin. If we know both the spectral shape and the energy-dependent detector resolution, this effect can be accounted for by integrating the contribution to the $i^{th}$ bin from each point in the spectrum. This type of analysis was done analytically in the previous $^{3}$H results~\cite{lux_tritium,attila}. However, the $S2$ tails described in section~\ref{sec:s2tails} make the analytic calculations unwieldy, so in this article we perform the integration numerically using Monte Carlo (MC) data.

In order to estimate these smearing corrections, the charge and light yield is initially taken from the NEST model~\cite{nest1,nest2,nest3}, and an initial set of MC $S1$ and $S2$ data is generated. This data~set is used to make preliminary measurements of $L_y$ and $Q_y$, after which the MC data are regenerated using these newly measured yields. This new set of MC data is used to re-measure the smearing corrections, giving us the final measurements of $L_y$ and $Q_y$.

\subsection{Combined Energy Model}\label{sec:combE}
We adapt the combined energy model for ER events~\cite{nest1,aprile_doke_LXe}, which relies on a simplified Platzman equation~\cite{platzman}:
\begin{equation}\label{eq:combe1} 
E_{ER}=W(n_*+n_i),
\end{equation}
where $n_*$ is the initial number of excitons generated by an event, and $n_i$ is the initial number of ions prior to recombination. For ER events, the work function, $W$, has been measured to be $13.7 \ \pm0.2$ eV/quantum~\cite{dahl}. Recombination converts some ions into excitons so that the observable number of photons ($n_{\gamma}$) and electrons ($n_{e}$) is given by:
\begin{equation}\label{eq:combe2}
\begin{split}
n_{\gamma}&=n_*+rn_i\\
&=(\alpha+r)n_i\\
n_e&=(1-r)n_i,
\end{split}
\end{equation}
where the exciton-ion ratio $\alpha \equiv n_{*}/n_{i}$, has been measured to be about 0.06-0.20~\cite{doke2002,attila} for ER events and is typically assumed to be constant with energy. For ER events, we assume a constant value of $\alpha=0.18$, which is taken from Ref.~\cite{attila}. For each event, the number of ions that recombine, $R$, is randomly distributed about an expected value that is equal to the mean recombination probability, $r$, times the number of ions:
\begin{equation}
    \langle R \rangle = r N_i .
\end{equation}
The mean recombination probability depends on both the energy deposited and the applied electric field.

We model this process using a modified version of the NEST simulation software~\cite{nest1,nest2,nest3}. The total number of quanta in a simulated event ($N_q$) is equal to the event energy, divided by the work function. The apportionment of these quanta into exitons and ions ($N_*$ and $N_i$) is treated as a binomial process, where the probability that a quantum is an ion is equal to $\frac{1}{1+\alpha}$. Recombination is modeled by drawing the number of electrons from a normal distribution with mean equal to $(1-r) \cdot N_i$ and standard deviation equal to $\sigma_R$, where $r$ and $\sigma_R$ depend on the energy and field of the simulated event. The number of photons ($N_{\gamma}$) is then taken to be the number of quanta, minus the number of electrons.  

\subsection{Measuring Average Charge and Light Collection Efficiencies}
Reconstructed energy is an observable quantity that fluctuates around the true energy deposited in the LXe during an event. In terms of the observable $S1$ and $S2$ signals, the reconstructed energy of an event is given by:
\begin{equation}\label{eq:combe3}
E_{rec}=W(\frac{S1}{g_1}+\frac{S2}{g_2}),
\end{equation}
where $g_1$ and $g_2$ are average gain factors. These values are used to convert from $S1$ to $n_{\gamma}$ and $S2$ to $n_{e}$, accounting for the total efficiency of the detector. Equation~\ref{eq:combe3} also provides a useful tool in measuring the efficiency factors, $g_1$ and $g_2$, through a method introduced by Doke~\textit{et al.} in Ref.~\cite{doke2002}. 

For a set of ER events with a constant energy $E$, Equation~\ref{eq:combe3} can be used to write:
\begin{equation}
\left(\frac{W\overline{S1}}{E}\right)=-\frac{g_1}{g_2}\left(\frac{W\overline{S2}}{E}\right)+g_1,
\end{equation}
where $\overline{S1}$ and $\overline{S2}$ are the average $S1$ and $S2$ signal observed. This linear expression may then be plotted in the usual way such that $y=\left(\frac{W\overline{S1}}{E}\right)$ and $x=\left(\frac{W\overline{S2}}{E}\right)$ are both in terms of either measurable quantities or known constants. The efficiency factors, $g_1$ and $g_2$, can then be obtained by fitting a line through a set of (x,y) values measured at different energies and fields.

\begin{figure}
\includegraphics[width=\linewidth]{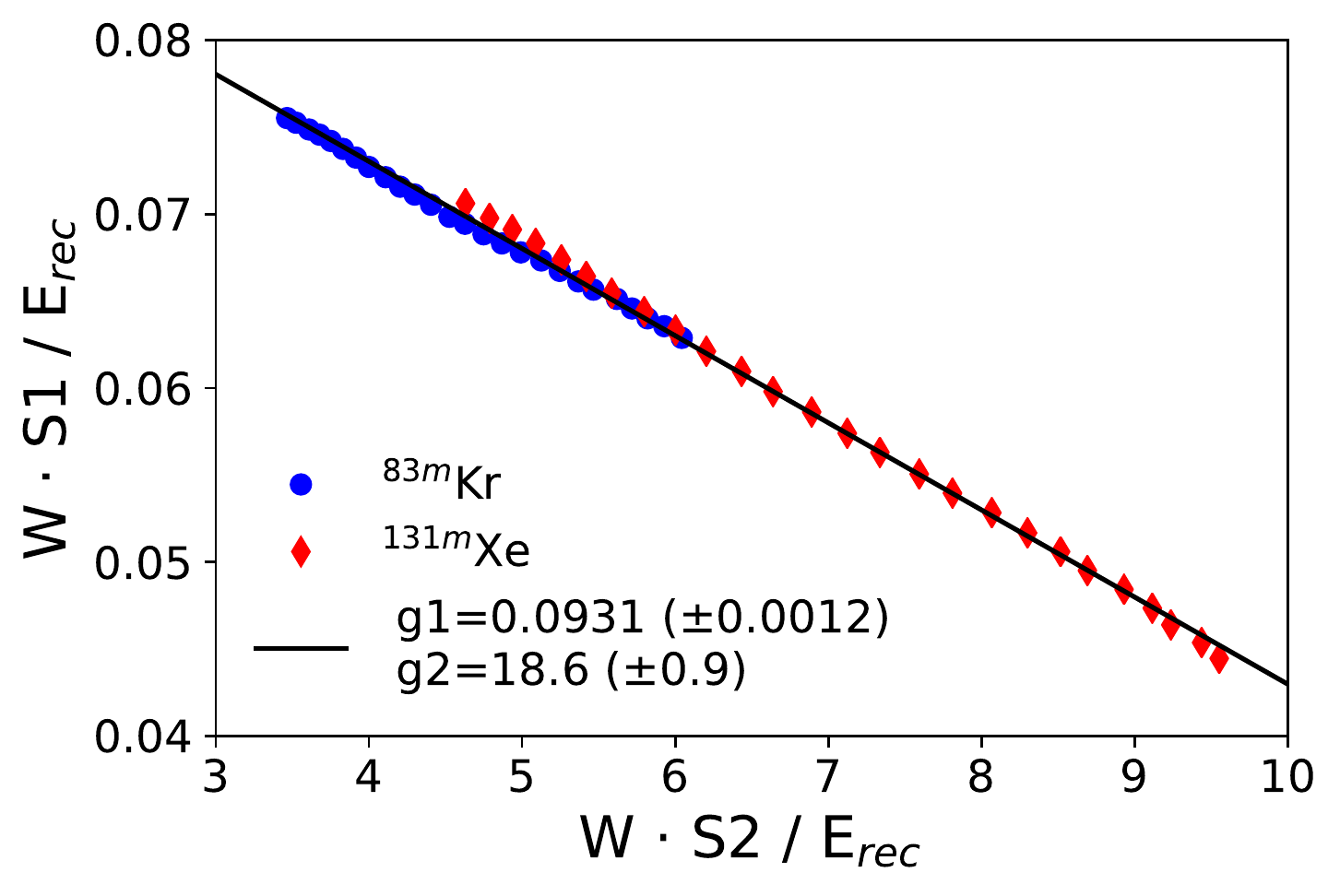}
\caption{Doke-style plot of post-WS $^{83m}$Kr and $^{131m}$Xe data. The uncertainties on the $g_1$ and $g_2$ values include systematic variation in drift time, and were calculated as described in the text. The $g_1$ and $g_2$ values are highly correlated, with a Pearson correlation coefficient of $-$0.90.}
 \label{fig:Doke_plot}
\end{figure}

For this analysis, the LUX post-WS values of $g_1$ and $g_2$ are measured by dividing the $^{83m}$Kr and $^{131m}$Xe data into separate drift-time regions and plotting the average $S1$ and $S2$ values in each region. The results of this analysis are shown in Figure~\ref{fig:Doke_plot}. To test for remaining position dependence in $g_1$ and $g_2$, we also calculate a two-point Doke plot using the $^{83m}$Kr and $^{131m}$Xe values from each drift-time region. The systematic uncertainty is taken to be the standard deviation of these two-point values. 

The values of $g_1$ and $g_2$ we measure are $0.0931~(\pm0.0012)$, and $18.6~(\pm0.9)$, respectively. The uncertainties of these values are dominated by the systematic deviation described above. These are broadly consistent with the values found in Ref.~\cite{lux_2017}, although our measured $g_1$ is about 3-$\sigma$ below the lowest value found there. This discrepancy is likely due to a continued decrease in light collection efficiency over the three months between the end of WS2014-2016 and the beginning of the injection campaign.

\subsection{Empirical Model of $S2$ Tail Pathology}\label{sec:s2tails}
\begin{figure}
\includegraphics[width=\linewidth]{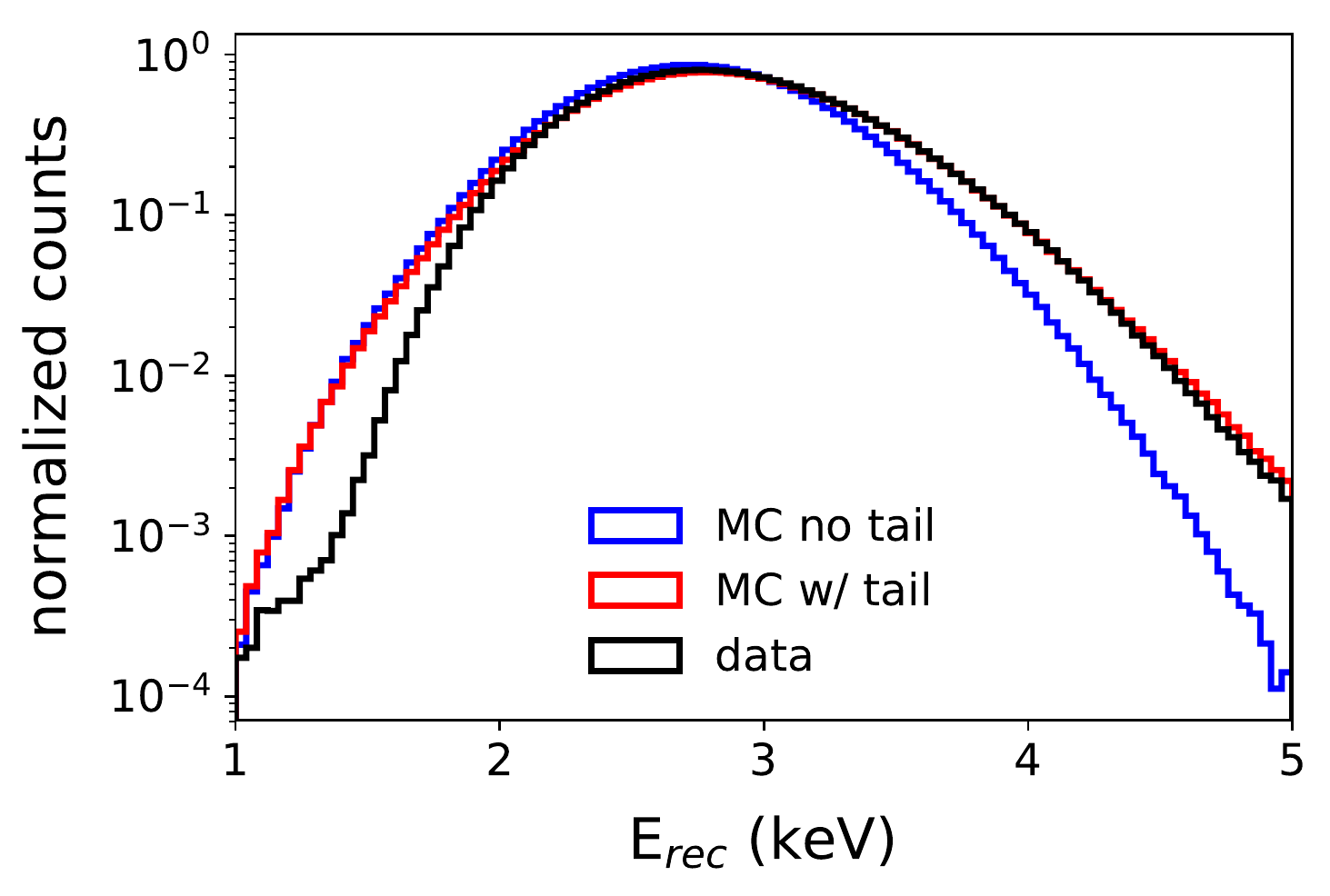}
\caption{Comparison of measured $^{37}$Ar energy spectrum (black) versus two simulated spectra; one with the $S2$ tails modeled (red) and one without (blue). The data shown are taken from the full WIMP-search fiducial volume corresponding to electron drift times of 40~to~300~microseconds.}
 \label{fig:Ar_spec}
\end{figure}

\begin{figure}
\includegraphics[width=\linewidth]{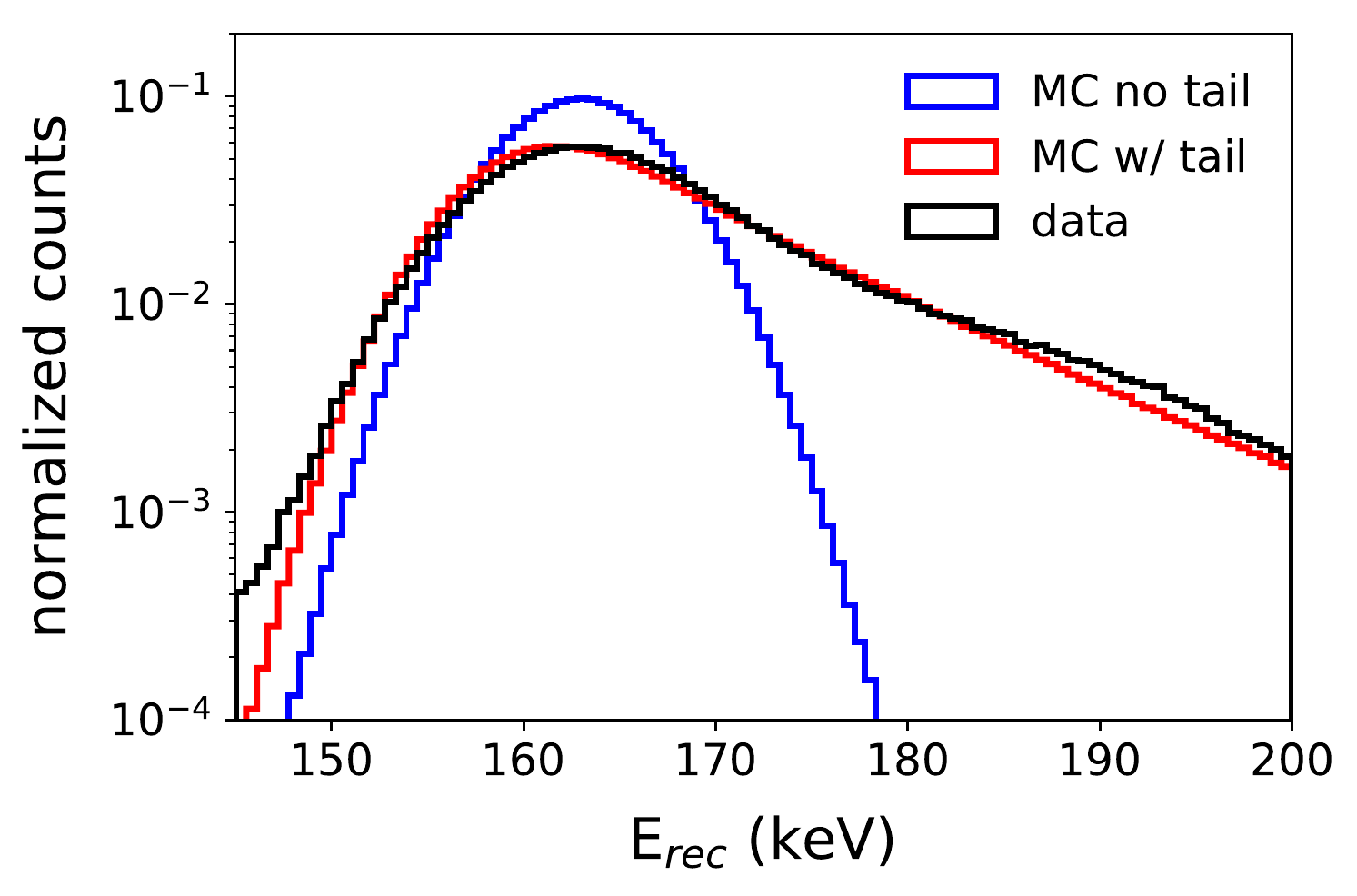}
\caption{Comparison of measured $^{131m}$Xe energy spectrum (black) versus two simulated spectra; one with the $S2$ tails modeled (red) and one without (blue). The data shown are taken from the full WIMP-search fiducial volume corresponding to electron drift times of 40~to~300~microseconds. }
 \label{fig:Xe_spec}
\end{figure}

Figures~\ref{fig:Ar_spec}~and~\ref{fig:Xe_spec} show the spectra of the $^{37}$Ar and $^{131m}$Xe decays measured during the post-WS calibration campaign. These combined energy spectra have clear non-Gaussian tails toward high energy. The tails in the energy spectra stem from underlying tails in the individual $S2$ spectra, which result from a pathological effect in the $S2$ signals. These $S2$ tails are much more pronounced in the WS2014-2016 and the post-WS data than in WS2013 data. The exact pathology is unknown, but there is some evidence that the tails are caused by either photoionization of impurities or ``electron trains''. An electron train occurs when electrons from a previous large event fail to be immediately extracted from the liquid surface and instead are emitted into the gas over a millisecond time scale.

To correctly account for smearing effects on the $^{14}$C and $^3$H spectra, we use an empirical model of the $S2$ tails. We begin with simulated $S1$ and $S2$ areas from MC events generated without tails, assuming Gaussian detector resolution. The effect of the $S2$ tails is modeled by assigning additional $S2$ area to a fraction of the simulated events. The additional tail area for a chosen event is drawn from an exponential distribution whose mean is proportional to uncorrected $S2$ size. This model of the $S2$ tails is generated using three steps and three fitting parameters, which are assumed to be independent of position, energy, and field. First, a ``true'' $g_2$ value ($g_{2,true}$) is used to generate an initial set of tail-less MC events. The value of $g_{2,true}$ is less than the one measured above, since the observed value of $g_2$ includes both the ``true'' $S2$ area, as well as the tail area. Next, a fraction of events, $R$, is selected to be assigned additional tail area. Finally, for each of the selected events, a random number of tail electrons is drawn from an exponential distribution with a mean of $b \cdot n_{e,LS}$, where $b$ is a fitting parameter, and $n_{e,LS}$ is the number of simulated electrons that reach the liquid surface. The parameters are tuned in order to reproduce the energy spectrum of the $^{37}$Ar and $^{131m}$Xe data. The best fit values of $b$, $R$, and $g_{2,true}$ are found to be $0.112~(\pm 0.003)$, $0.73~(\pm 0.02)$, and $17.60~(\pm 0.05)$, respectively. 

\begin{figure}
\includegraphics[width=\linewidth]{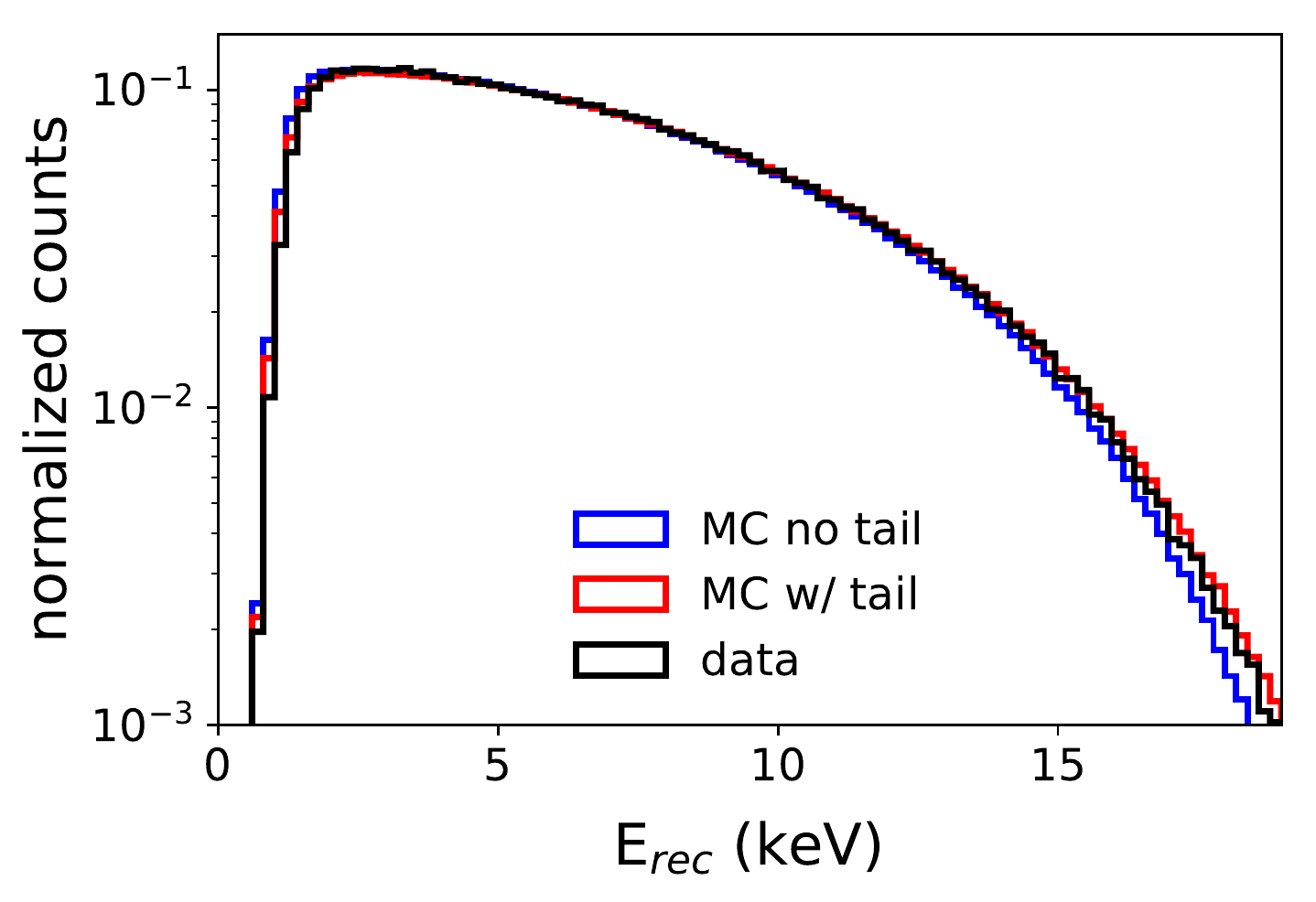}
\caption{Comparison of measured $^{3}$H energy spectrum (black) versus two simulated spectra; one with the $S2$ tails modeled (red) and one without (blue). The data shown are taken from the full WIMP-search fiducial volume corresponding to electron drift times of 40~to~300~microseconds.}
 \label{fig:H3_spec}
\end{figure}

\begin{figure}
\includegraphics[width=\linewidth]{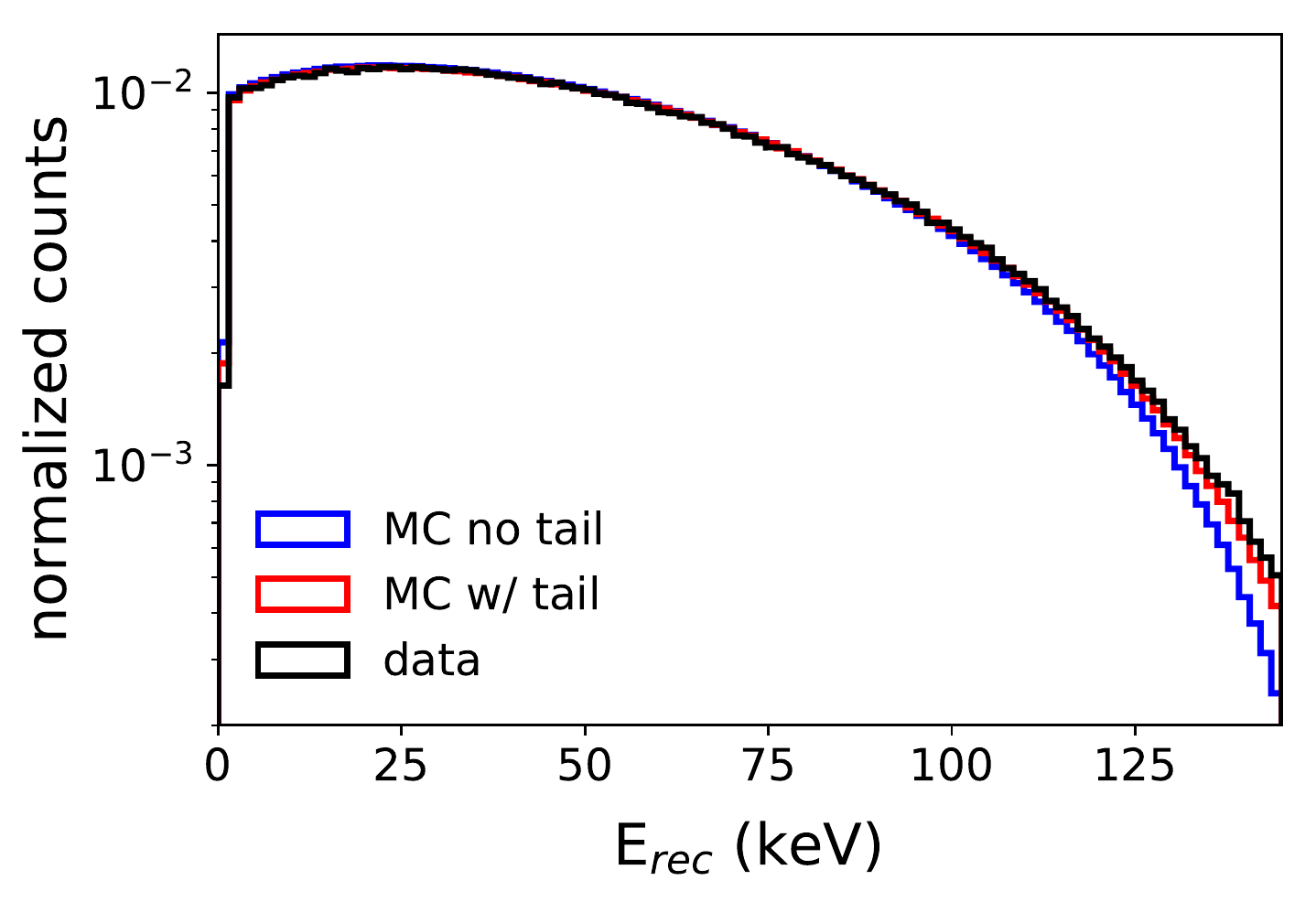}
\caption{Comparison of measured $^{14}$C energy spectrum (black) versus two simulated spectra; one with the $S2$ tails modeled (red) and one without (blue). The data shown are taken from the full WIMP-search fiducial volume corresponding to electron drift times of 40~to~300~microseconds.}
 \label{fig:C14_spec}
\end{figure}

Figures~\ref{fig:Ar_spec} and~\ref{fig:Xe_spec} show the best fit spectra for $^{37}$Ar and $^{131m}$Xe. Figures~\ref{fig:H3_spec} and~\ref{fig:C14_spec} show the best fit model applied to $^{3}$H and $^{14}$C, respectively. The agreement of all of the simulated spectra with data are significantly improved after the addition of the tail model. The $^{37}$Ar MC spectrum over-predicts the amplitude of the data at low energy ($<$2 keV); however, the same discrepancy is not observed in the $^{3}$H spectrum.

\subsection{Recombination Fluctuations}\label{sec:sigr_mod}
The fluctuations in the recombination fraction, $\sigma_R$, are known to deviate significantly from those of a binomial process~\cite{dahl,attila,lux_tritium}. In Ref.~\cite{lux_tritium} we reported that the fluctuations are approximately linear in $n_i$ with a slope of about 0.067~quanta~per~ion. We do not attempt to obtain an absolute measurement of $\sigma_R$ using the post-WS data because the $S2$ tails are correlated with the recombination fluctuations in nontrivial ways. The correlation makes it impractical to separate detector resolution and recombination fluctuations as was done for the WS2013 $^{3}$H data. We instead apply an adjustment to the linear model and compare the resulting MC spectrum to data. We find the data are best described by a Gaussian adjustment to the linear model:
\begin{equation}\label{eq:sigrmod}
\begin{split}
\sigma_{R}(r)^2=r&(1-r)\cdot n_i +\\
&\left(F_0\exp \left(\frac{-(r-F_1)^2}{2F_2^2}\right)\right)^2n_i^2,
\end{split}
\end{equation}
where $F_0$, $F_1$ and $F_2$ are the constant fitting parameters:
\begin{equation}\label{eq:sigrbestfit}
\begin{split}
F_0=0.075 \pm 0.005\\
F_1=0.413 \pm 0.024\\
F_2=0.243 \pm 0.024 .
\end{split}
\end{equation}
These parameters were optimized using post-WS $^3$H and $^{14}$C data using using a grid search method described in section~5.5.3 of Ref.~\cite{thesis}. A set of measurements of $\sigma_R$ from Ref.~\cite{dahl} were also used to help constrain the low-recombination side of the model. When $N_i$ is large and $r$ is close to $F_1$, Equation~\ref{eq:sigrmod} reduces to a linear model with a slope of $0.075  (\pm 0.005)$~quanta~per~ion. The first term on the right side of Equation~\ref{eq:sigrmod} mimics a binomial variance and prevents $\sigma_R$ from going to zero at extreme values or r. In our best fit model, the binomial term is negligible across the range of energies and fields tested.

\begin{figure}[h!]
\includegraphics[width=\linewidth]{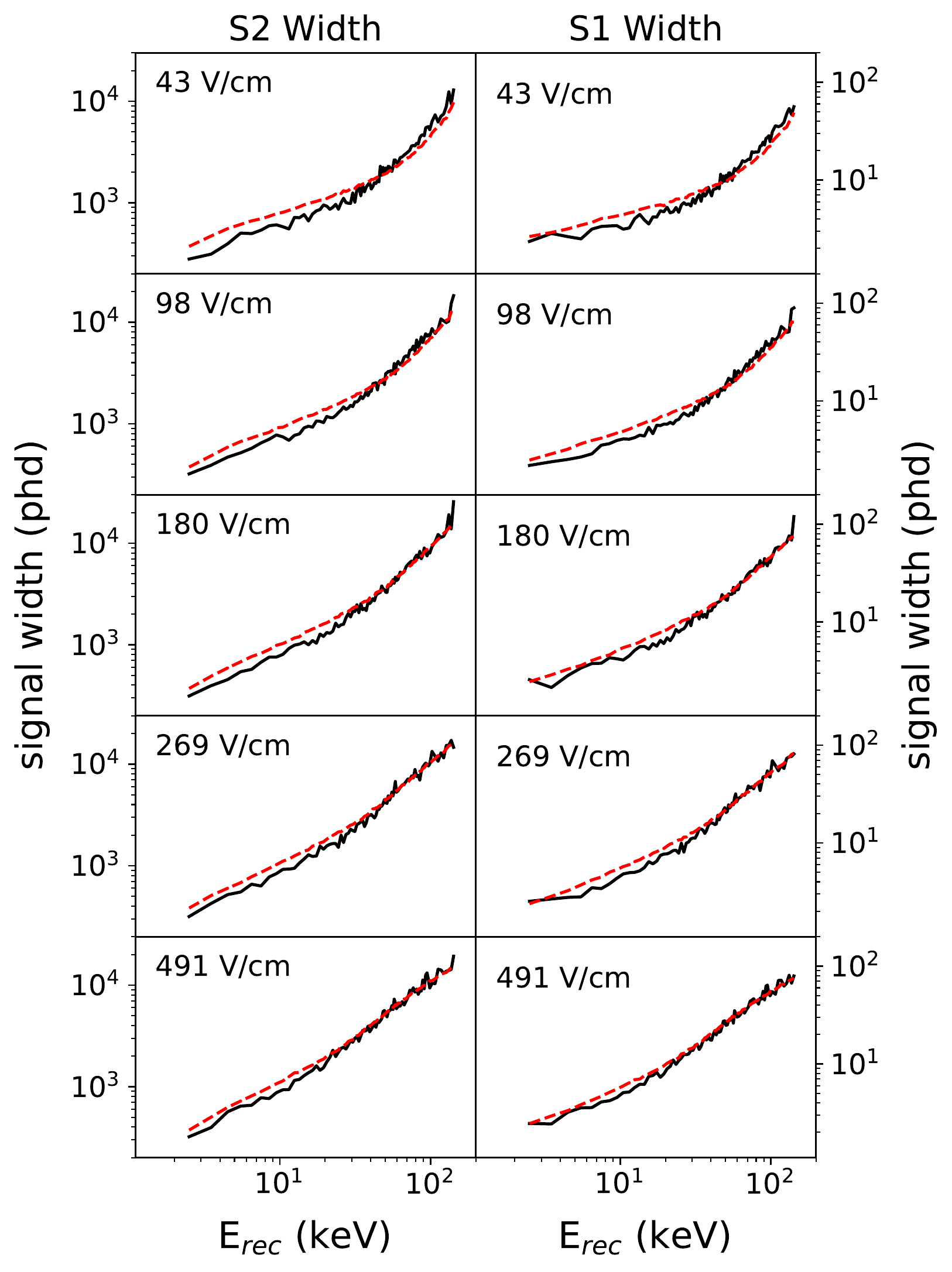}[h]
\caption{This figure shows the best-fit Gaussian width of the $^{14}$C $S1$ (right) and $S2$ (left) bands for a selection of electric field bins. In these plots, the red dashed lines indicate simulated widths using the model described in sections~\ref{sec:s2tails} and~\ref{sec:sigr_mod}, while the black lines indicate the widths observed in data.}
 \label{fig:C14_sigR_sim}
\end{figure}

\begin{figure}[h!]
\includegraphics[width=\linewidth]{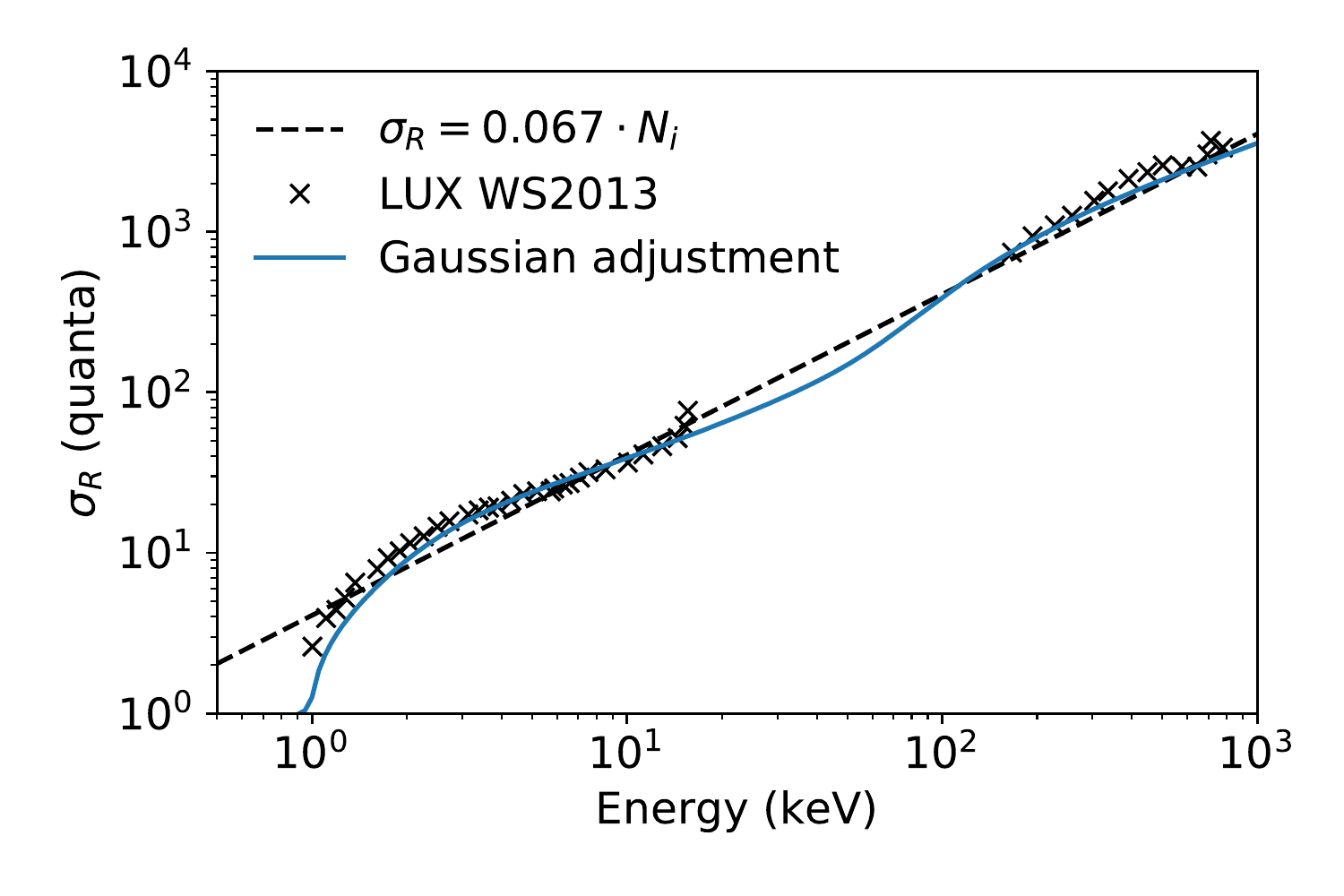}
\caption{Comparison of the recombination model developed in section~\ref{sec:sigr_mod} (blue line) to the model from~\cite{lux_tritium} (black dashed line). The black markers indicate the $\sigma_R$ data presented in~\cite{attila}.}
 \label{fig:sigR_run03_comp}
\end{figure}

The width of the MC spectra are compared to data in Figure~\ref{fig:C14_sigR_sim}. We find that by adding the $S2$ tail model described in section~\ref{sec:s2tails} and a model of recombination fluctuations that follows Equation~\ref{eq:sigrmod} to our simulation, we are able to reproduce the widths of the $S1$ and $S2$ spectra for $^{14}$C $\beta$-decay events across all of the electric fields tested. Figure~\ref{fig:sigR_run03_comp} shows a comparison to the $\sigma_R$ measurements from WS2013. We find the new model matches data better than the linear model. The upward kink in the WS2013 measurements at 16~keV is due to an error that will be described in section~\ref{sec:discussion}. 

There is still tension remaining between simulated and measured widths with this new model included. It may be that this is due to an underlying field dependence in the recombination fluctuations that has not been unaccounted for. This would be a third-order effect in our measurements of the yields, and we are able to reproduce the measured $^3$H and $^{14}$C spectra without accounting for this possible field dependence. We therefore elect to proceed using the model as described above.

\section{Results and Discussion}

\subsection{Photon-Electron Fraction}
Assuming that for ER the total number of generated quanta is fixed, as described in section~\ref{sec:combE}, we can reduce $L_y$, $Q_y$, and $r$ to a single quantity:
\begin{equation}
\rho \equiv \frac{n_{\gamma}}{n_e}.
\end{equation}
Using the relations:
\begin{equation}
L_y\equiv \frac{n_{\gamma}}{E}, \ Q_y\equiv \frac{n_{e}}{E},\text{ and } 
L_y+Q_y=\frac{1}{W},
\end{equation}
we can reconstruct the individual yields from $\rho$:
\begin{equation}\label{eq:translate_rho1}
L_y=\frac{1}{W}\frac{\rho}{1+\rho} \text{  and  } Q_y=\frac{1}{W}\frac{1}{1+\rho}.
\end{equation}
Further, using the relations:
\begin{equation}
n_Q=(1+\alpha)n_i=(1+\rho)n_e \text{  and  } 
r=\frac{n_i-n_e}{n_i},
\end{equation}
where $n_Q$ is the total number of quanta, we can reconstruct the average recombination probability:
\begin{equation}\label{eq:translate_rho2}
r=\frac{\rho-\alpha}{1+\rho}.
\end{equation}

\begin{figure*}[h]
\includegraphics[width=\linewidth]{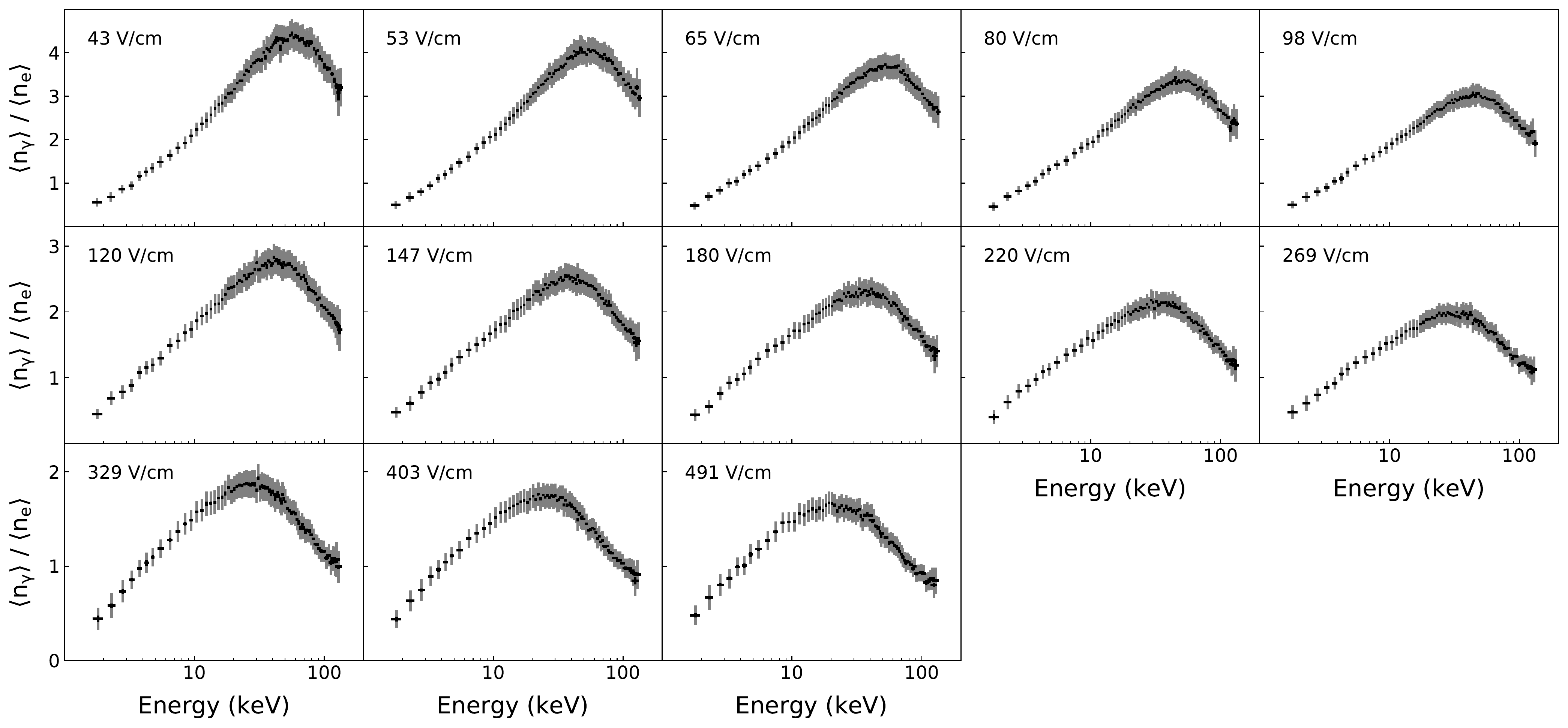}
\caption{Measurements of $\langle \rho \rangle$ for post-WS $^{14}$C data in the specified electric field bins. The dark error bars show the statistical uncertainty, and the light error bars show the systematic plus statistical uncertainty.}
 \label{fig:C14_rho_final_z}
\end{figure*}

\begin{figure*}[h!]
\includegraphics[width=\textwidth]{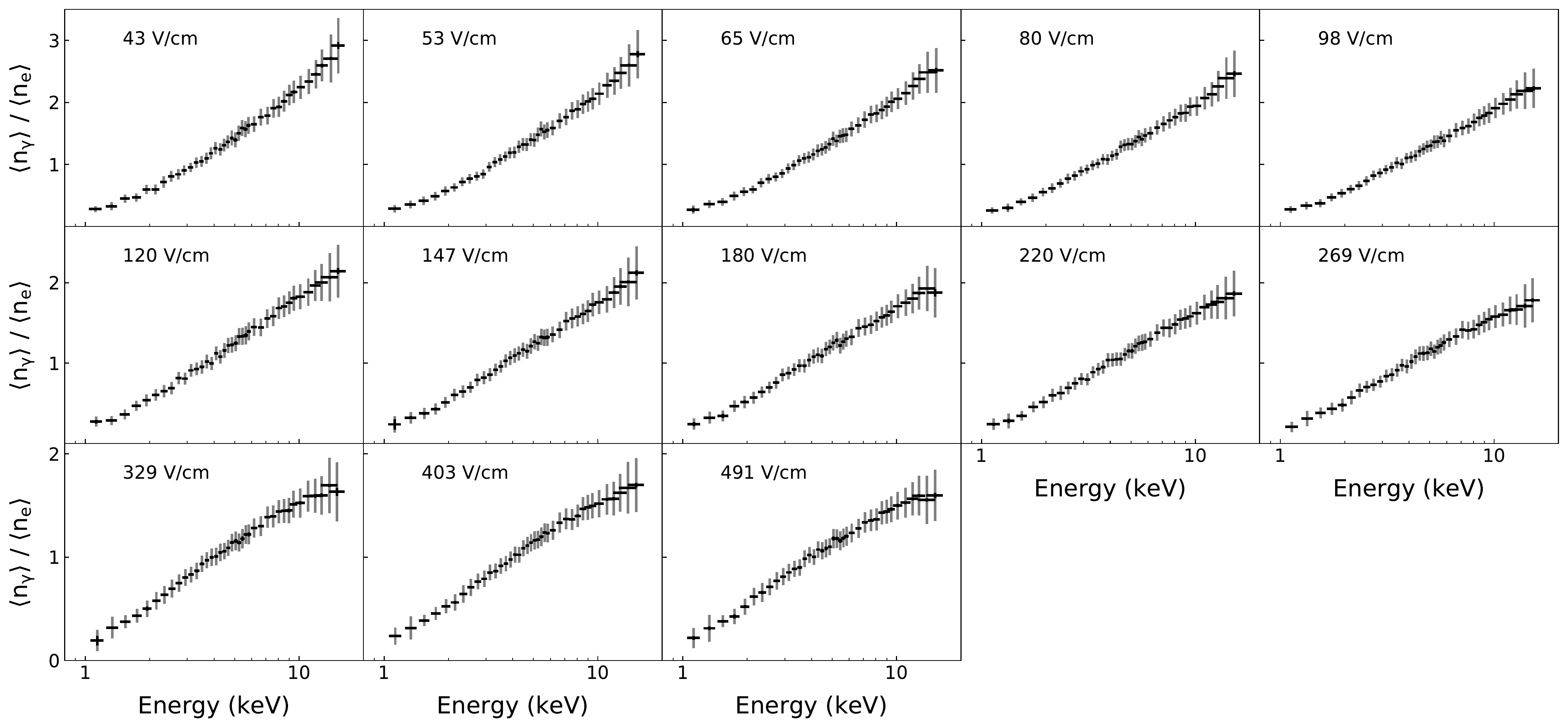}
\caption{Measurements of $\langle \rho \rangle$ for post-WS $^{3}$H data in the specified electric field bins. The dark error bars show the statistical uncertainty, and the light error bars show the systematic plus statistical uncertainty.}
 \label{fig:H3_rho_final_z}
\end{figure*}

Here we report the measured results of $\rho\equiv n_{\gamma}/n_e$. Figure~\ref{fig:C14_rho_final_z} shows the results for post-WS $^{14}$C, and figure~\ref{fig:H3_rho_final_z} shows the results for post-WS $^{3}$H. The measurements of $n_{\gamma}$ and $n_e$, along with the reconstructed energy have been numerically de-smeared following the procedure laid out in section~\ref{sec:desmear}, using the model described in sections~\ref{sec:s2tails} and~\ref{sec:sigr_mod}. The sizes of these smearing corrections are taken as systematic uncertainty on their respective measurements. The uncertainties in $g_1$ and $g_2$ are also included in the systematic error. The systematic uncertainties are combined in quadrature and are shown as the light gray error bars in figures~\ref{fig:C14_rho_final_z} and~\ref{fig:H3_rho_final_z}. The statistical fitting uncertainty and the uncertainty due to bin width are shown as the black error bars.

%\section{Empirical Model of Q_y}\label{sec:betamodel}
%The wide range of energies and fields we probed by the post-WS beta calibrations allows us to build an empirical model of the charge yield. We found that a combination of two asymmetric sigmoids is able to well characterize the measurements at all fields and energies:
%\begin{equation}
%Q_y(E,\mathcal{E})=m_1+\frac{m_2-m_1}{(1+(E/m_3)^{m_4})^{m_9}}+m_5+\frac{0-m_5}{(1+(E/m_7)^{m_8})^{m_{10}}},
%\end{equation}
%where $\mathcal{E}$ is the electric field in V/cm, and $m_1$ through $m_10$ are fitting parameters. Additionally, the $m_1$ and $m_7$ parameters are allowed to vary with electric field.

%\begin{figure}
%\includegraphics[width=\linewidth]{model_field_dep}
%\caption{}
% \label{fig:model_field_dep}
%\end{figure}

\subsection{Discussion}\label{sec:discussion}
\begin{figure}
\includegraphics[width=\linewidth]{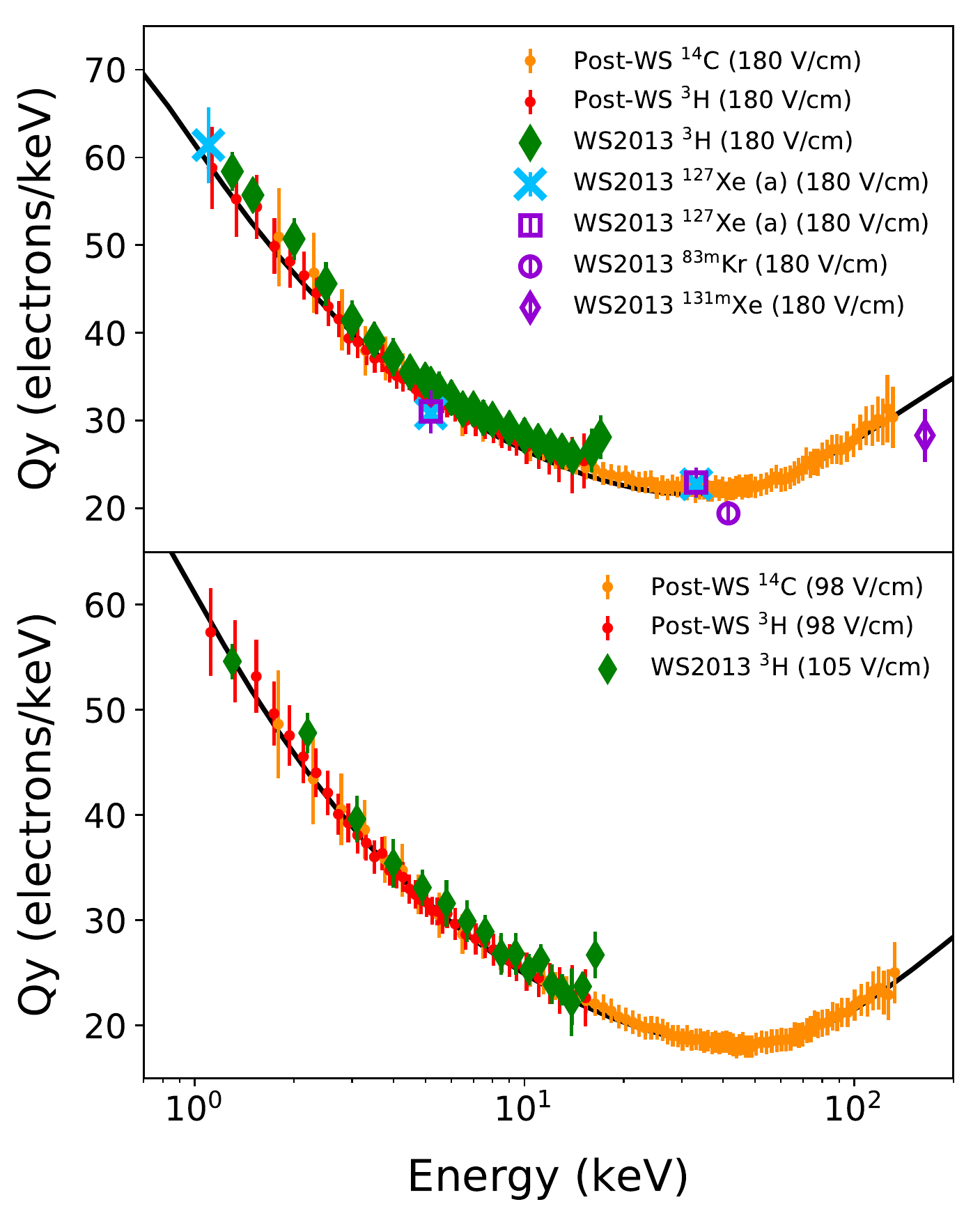}
\caption{Comparison of LUX post-WS $Q_y$ measurements using $^{3}$H (red) and $^{14}$C (orange) to WS2013 measurements, which were taken at 105~and~180~V/cm. The green diamonds show the WS2013 $^{3}$H measurements~\cite{lux_tritium}, and the blue X's and open magenta squares indicate WS2013 measurements of $Q_y$ from $^{127}$Xe electron capture~\cite{DQyields,Evanyields}. The open circles and diamonds indicate WS2013 measurements of $Q_y$ from the $^{83m}$Kr and $^{131m}$Xe decays, respectively~\cite{Evanyields}. The black line shows the final model used to generate the smearing corrections~\cite{thesis}.}
 \label{fig:QY_run03_comp}
\end{figure}

Our results are in good agreement with previous measurements from WS2013, as is shown in figure~\ref{fig:QY_run03_comp}. In this figure we compare measurements of $Q_y$ from WS2013 $^{3}$H~\cite{lux_tritium} and from the $^{127}$Xe electron capture~\cite{DQyields,Evanyields} with those from this work at similar electric fields. We find that the measurements agree within systematic error. 
When comparing our measurements of $Q_y$ from interactions of $\beta$'s in LXe to those from the $^{83m}$Kr and $^{131m}$Xe decays we find a disagreement of about 2-$\sigma$~\cite{Evanyields}. This is likely due to a difference in yields between $\beta$-decay interactions and those involving composite decays (such as $^{83m}$Kr) or photo-absorption~\cite{nest2}.


It should be noted that the results we present here are in disagreement with our previous $^3$H yields and recombination measurements from Ref.~\cite{lux_tritium} above 16~keV. In the previous work, an error in the implementation of the energy smearing correction resulted in some of the data being over-corrected. The error has only a minimal effect for most of the results reported in Ref.~\cite{lux_tritium}, but it is manifest at the endpoint of the tritium spectrum as a kink in the yields. A detailed discussion of this error can be found in section~5.3.2 of Ref.~\cite{thesis}.

\section{Summary}
We have presented improved measurements of the response of liquid xenon to $\beta$-decays in the LUX detector, which were taken after WS2014-2016 was completed. We describe the various sources used, along with the time-line and the respective activities of the calibrations. We use the $^{83m}$Kr and $^{131m}$Xe lines to measure the average $g_1$ and $g_2$ efficiency factors and to characterize the positional variation thereof. 

The $^{37}$Ar and $^{131m}$Xe calibration data are used to alter the existing model of detector resolution to account for the $S2$ tail pathology in the $S2$ spectra. We used this updated model to numerically calculate the effect of smearing on the $^{3}$H and $^{14}$C $\beta$-decay spectra. We also found it necessary to update the empirical model of recombination fluctuations presented in~\cite{lux_tritium} to better match the data above 20~keV.

We present measurements of the photon-to-electron ratio of $\beta$ events in liquid xenon from $^{3}$H and $^{14}$C. These measurements can be used to calculate the charge yield, light yield, and recombination probability over a wide range of electric fields and energies. This is the most extensive dataset of the quantities for $\beta$-decay in liquid xenon, and are directly relevant for understanding the dominant background in future dark matter experiments.

\begin{acknowledgments}

This work was partially supported by the U.S. Department of Energy (DOE) under award numbers DE-FG02-08ER41549, DE-FG02-91ER40688, DE-FG02-95ER40917, DE-FG02-91ER40674, DE-NA0000979, DE-FG02-11ER41738, DE-SC0006605, DE-AC02-05CH11231, DE-AC52-07NA27344, DE-SC0019066, and DE-FG01-91ER40618; the U.S. National Science Foundation under award numbers PHYS-0750671, PHY-0801536, PHY-1004661, PHY-1102470, PHY-1003660, PHY-1312561, PHY-1347449; the Research Corporation grant RA0350; the Center for Ultra-low Background Experiments in the Dakotas (CUBED); and the South Dakota School of Mines and Technology (SDSMT). LIP-Coimbra acknowledges funding from Funda\c{c}\~{a}o para a Ci\^{e}ncia e a Tecnologia (FCT)   through the project-grant PTDC/FIS-PAR/28567/2017. Imperial College and Brown University thank the UK Royal Society for travel funds under the International Exchange Scheme (IE120804). The UK groups acknowledge institutional support from Imperial College London, University College London and Edinburgh University, and from the Science \& Technology Facilities Council for PhD studentship ST/K502042/1 (AB). The University of Edinburgh is a charitable body, registered in Scotland, with registration number SC005336. This research was conducted using computational resources and services at the Center for Computation and Visualization, Brown University.

We gratefully acknowledge the logistical and technical support and the access to laboratory infrastructure provided to us by the Sanford Underground Research Facility (SURF) and its personnel at Lead, South Dakota. SURF was developed by the South Dakota Science and Technology Authority, with an important philanthropic donation from T. Denny Sanford, and is operated by Lawrence Berkeley National Laboratory for the Department of Energy, Office of High Energy Physics.

\end{acknowledgments}

%double blank page so that the table doesn't spill into the bibliography
%\newpage
%\thispagestyle{empty}
%\mbox{}
%\newpage
%\thispagestyle{empty}
%\mbox{}

% \bibliography{C14_paper}

\begin{thebibliography}{32}%
\makeatletter
\providecommand \@ifxundefined [1]{%
 \@ifx{#1\undefined}
}%
\providecommand \@ifnum [1]{%
 \ifnum #1\expandafter \@firstoftwo
 \else \expandafter \@secondoftwo
 \fi
}%
\providecommand \@ifx [1]{%
 \ifx #1\expandafter \@firstoftwo
 \else \expandafter \@secondoftwo
 \fi
}%
\providecommand \natexlab [1]{#1}%
\providecommand \enquote  [1]{``#1''}%
\providecommand \bibnamefont  [1]{#1}%
\providecommand \bibfnamefont [1]{#1}%
\providecommand \citenamefont [1]{#1}%
\providecommand \href@noop [0]{\@secondoftwo}%
\providecommand \href [0]{\begingroup \@sanitize@url \@href}%
\providecommand \@href[1]{\@@startlink{#1}\@@href}%
\providecommand \@@href[1]{\endgroup#1\@@endlink}%
\providecommand \@sanitize@url [0]{\catcode `\\12\catcode `\$12\catcode
  `\&12\catcode `\#12\catcode `\^12\catcode `\_12\catcode `\%12\relax}%
\providecommand \@@startlink[1]{}%
\providecommand \@@endlink[0]{}%
\providecommand \url  [0]{\begingroup\@sanitize@url \@url }%
\providecommand \@url [1]{\endgroup\@href {#1}{\urlprefix }}%
\providecommand \urlprefix  [0]{URL }%
\providecommand \Eprint [0]{\href }%
\providecommand \doibase [0]{http://dx.doi.org/}%
\providecommand \selectlanguage [0]{\@gobble}%
\providecommand \bibinfo  [0]{\@secondoftwo}%
\providecommand \bibfield  [0]{\@secondoftwo}%
\providecommand \translation [1]{[#1]}%
\providecommand \BibitemOpen [0]{}%
\providecommand \bibitemStop [0]{}%
\providecommand \bibitemNoStop [0]{.\EOS\space}%
\providecommand \EOS [0]{\spacefactor3000\relax}%
\providecommand \BibitemShut  [1]{\csname bibitem#1\endcsname}%
\let\auto@bib@innerbib\@empty
%</preamble>
\bibitem [{\citenamefont {Akerib}\ \emph {et~al.}(2013)\citenamefont {Akerib}
  \emph {et~al.}}]{lux_2012}%
  \BibitemOpen
  \bibfield  {author} {\bibinfo {author} {\bibfnamefont {D.~S.}\ \bibnamefont
  {Akerib}} \emph {et~al.} (\bibinfo {collaboration} {LUX Collaboration}),\
  }\href {\doibase 10.1016/j.nima.2012.11.135} {\bibfield  {journal} {\bibinfo
  {journal} {Nucl. Instrum. Meth.}\ }\textbf {\bibinfo {volume} {A704}},\
  \bibinfo {pages} {111} (\bibinfo {year} {2013})},\ \Eprint
  {http://arxiv.org/abs/1211.3788} {arXiv:1211.3788 [physics.ins-det]}
  \BibitemShut {NoStop}%
%%CITATION = ARXIV:1211.3788;%%
\bibitem [{\citenamefont {Akerib}\ \emph {et~al.}(2014)\citenamefont {Akerib}
  \emph {et~al.}}]{lux_2014}%
  \BibitemOpen
  \bibfield  {author} {\bibinfo {author} {\bibfnamefont {D.~S.}\ \bibnamefont
  {Akerib}} \emph {et~al.} (\bibinfo {collaboration} {LUX Collaboration}),\
  }\href {\doibase 10.1103/PhysRevLett.112.091303} {\bibfield  {journal}
  {\bibinfo  {journal} {Phys. Rev. Lett.}\ }\textbf {\bibinfo {volume} {112}},\
  \bibinfo {pages} {091303} (\bibinfo {year} {2014})}\BibitemShut {NoStop}%
\bibitem [{\citenamefont {Akerib}\ \emph
  {et~al.}(2016{\natexlab{a}})\citenamefont {Akerib} \emph
  {et~al.}}]{lux_2016}%
  \BibitemOpen
  \bibfield  {author} {\bibinfo {author} {\bibfnamefont {D.~S.}\ \bibnamefont
  {Akerib}} \emph {et~al.} (\bibinfo {collaboration} {LUX Collaboration}),\
  }\href {\doibase 10.1103/PhysRevLett.116.161301} {\bibfield  {journal}
  {\bibinfo  {journal} {Phys. Rev. Lett.}\ }\textbf {\bibinfo {volume} {116}},\
  \bibinfo {pages} {161301} (\bibinfo {year} {2016}{\natexlab{a}})}\BibitemShut
  {NoStop}%
\bibitem [{\citenamefont {Akerib}\ \emph
  {et~al.}(2017{\natexlab{a}})\citenamefont {Akerib} \emph
  {et~al.}}]{lux_2017}%
  \BibitemOpen
  \bibfield  {author} {\bibinfo {author} {\bibfnamefont {D.~S.}\ \bibnamefont
  {Akerib}} \emph {et~al.} (\bibinfo {collaboration} {LUX Collaboration}),\
  }\href {\doibase 10.1103/PhysRevLett.118.021303} {\bibfield  {journal}
  {\bibinfo  {journal} {Phys. Rev. Lett.}\ }\textbf {\bibinfo {volume} {118}},\
  \bibinfo {eid} {021303} (\bibinfo {year} {2017}{\natexlab{a}})},\ \Eprint
  {http://arxiv.org/abs/1608.07648} {arXiv:1608.07648} \BibitemShut {NoStop}%
\bibitem [{\citenamefont {Aprile}\ \emph {et~al.}(2018)\citenamefont {Aprile}
  \emph {et~al.}}]{xenon_1t}%
  \BibitemOpen
  \bibfield  {author} {\bibinfo {author} {\bibfnamefont {E.}~\bibnamefont
  {Aprile}} \emph {et~al.} (\bibinfo {collaboration} {XENON Collaboration 7}),\
  }\href {\doibase 10.1103/PhysRevLett.121.111302} {\bibfield  {journal}
  {\bibinfo  {journal} {Phys. Rev. Lett.}\ }\textbf {\bibinfo {volume} {121}},\
  \bibinfo {pages} {111302} (\bibinfo {year} {2018})}\BibitemShut {NoStop}%
\bibitem [{\citenamefont {Cui}\ \emph {et~al.}(2017)\citenamefont {Cui} \emph
  {et~al.}}]{pandax}%
  \BibitemOpen
  \bibfield  {author} {\bibinfo {author} {\bibfnamefont {X.}~\bibnamefont
  {Cui}} \emph {et~al.} (\bibinfo {collaboration} {PandaX-II Collaboration}),\
  }\href {\doibase 10.1103/PhysRevLett.119.181302} {\bibfield  {journal}
  {\bibinfo  {journal} {Phys. Rev. Lett.}\ }\textbf {\bibinfo {volume} {119}},\
  \bibinfo {pages} {181302} (\bibinfo {year} {2017})}\BibitemShut {NoStop}%
\bibitem [{\citenamefont {Szydagis}\ \emph {et~al.}(2013)\citenamefont
  {Szydagis}, \citenamefont {Fyhrie}, \citenamefont {Thorngren},\ and\
  \citenamefont {Tripathi}}]{nest2}%
  \BibitemOpen
  \bibfield  {author} {\bibinfo {author} {\bibfnamefont {M.}~\bibnamefont
  {Szydagis}}, \bibinfo {author} {\bibfnamefont {A.}~\bibnamefont {Fyhrie}},
  \bibinfo {author} {\bibfnamefont {D.}~\bibnamefont {Thorngren}}, \ and\
  \bibinfo {author} {\bibfnamefont {M.}~\bibnamefont {Tripathi}},\ }\href@noop
  {} {\bibfield  {journal} {\bibinfo  {journal} {Journal of Instrumentation}\
  }\textbf {\bibinfo {volume} {8}},\ \bibinfo {pages} {C10003} (\bibinfo {year}
  {2013})}\BibitemShut {NoStop}%
\bibitem [{\citenamefont {Mount}\ \emph {et~al.}(2017)\citenamefont {Mount}
  \emph {et~al.}}]{lz_tdr}%
  \BibitemOpen
  \bibfield  {author} {\bibinfo {author} {\bibfnamefont {B.~J.}\ \bibnamefont
  {Mount}} \emph {et~al.},\ }\href@noop {} {\  (\bibinfo {year} {2017})},\
  \Eprint {http://arxiv.org/abs/1703.09144} {arXiv:1703.09144
  [physics.ins-det]} \BibitemShut {NoStop}%
%%CITATION = ARXIV:1703.09144;%%
\bibitem [{\citenamefont {Akerib}\ \emph
  {et~al.}(2018{\natexlab{a}})\citenamefont {Akerib} \emph
  {et~al.}}]{lz_sensitivity}%
  \BibitemOpen
  \bibfield  {author} {\bibinfo {author} {\bibfnamefont {D.~S.}\ \bibnamefont
  {Akerib}} \emph {et~al.} (\bibinfo {collaboration} {LUX-ZEPLIN}),\
  }\href@noop {} {\  (\bibinfo {year} {2018}{\natexlab{a}})},\ \Eprint
  {http://arxiv.org/abs/1802.06039} {arXiv:1802.06039 [astro-ph.IM]}
  \BibitemShut {NoStop}%
%%CITATION = ARXIV:1802.06039;%%
\bibitem [{\citenamefont {Akerib}\ \emph
  {et~al.}(2016{\natexlab{b}})\citenamefont {Akerib} \emph
  {et~al.}}]{lux_tritium}%
  \BibitemOpen
  \bibfield  {author} {\bibinfo {author} {\bibfnamefont {D.~S.}\ \bibnamefont
  {Akerib}} \emph {et~al.} (\bibinfo {collaboration} {LUX Collaboration}),\
  }\href {\doibase 10.1103/PhysRevD.93.072009} {\bibfield  {journal} {\bibinfo
  {journal} {Phys. Rev.}\ }\textbf {\bibinfo {volume} {D93}},\ \bibinfo {pages}
  {072009} (\bibinfo {year} {2016}{\natexlab{b}})},\ \Eprint
  {http://arxiv.org/abs/1512.03133} {arXiv:1512.03133 [physics.ins-det]}
  \BibitemShut {NoStop}%
%%CITATION = ARXIV:1512.03133;%%
\bibitem [{\citenamefont {Akerib}\ \emph
  {et~al.}(2017{\natexlab{b}})\citenamefont {Akerib} \emph
  {et~al.}}]{DQyields}%
  \BibitemOpen
  \bibfield  {author} {\bibinfo {author} {\bibfnamefont {D.~S.}\ \bibnamefont
  {Akerib}} \emph {et~al.} (\bibinfo {collaboration} {LUX Collaboration}),\
  }\href@noop {} {\  (\bibinfo {year} {2017}{\natexlab{b}})},\ \Eprint
  {http://arxiv.org/abs/1709.00800} {arXiv:1709.00800 [physics.ins-det]}
  \BibitemShut {NoStop}%
%%CITATION = ARXIV:1709.00800;%%
\bibitem [{\citenamefont {Akerib}\ \emph
  {et~al.}(2017{\natexlab{c}})\citenamefont {Akerib} \emph
  {et~al.}}]{Evanyields}%
  \BibitemOpen
  \bibfield  {author} {\bibinfo {author} {\bibfnamefont {D.~S.}\ \bibnamefont
  {Akerib}} \emph {et~al.} (\bibinfo {collaboration} {LUX Collaboration}),\
  }\href {\doibase 10.1103/PhysRevD.95.012008} {\bibfield  {journal} {\bibinfo
  {journal} {Phys. Rev.}\ }\textbf {\bibinfo {volume} {D95}},\ \bibinfo {pages}
  {012008} (\bibinfo {year} {2017}{\natexlab{c}})},\ \Eprint
  {http://arxiv.org/abs/1610.02076} {arXiv:1610.02076 [physics.ins-det]}
  \BibitemShut {NoStop}%
%%CITATION = ARXIV:1610.02076;%%
\bibitem [{\citenamefont {Balajthy}(2018)}]{thesis}%
  \BibitemOpen
  \bibfield  {author} {\bibinfo {author} {\bibfnamefont {J.}~\bibnamefont
  {Balajthy}},\ }\emph {\bibinfo {title} {Purity Monitoring Techniques and
  Electronic Energy Deposition Properties in Liquid Xenon Time Projection
  Chambers}},\ \href {\doibase doi:10.13016/M2BG2HD6T} {Ph.D. thesis},\
  \bibinfo  {school} {Uniersity of Maryland} (\bibinfo {year}
  {2018})\BibitemShut {NoStop}%
\bibitem [{\citenamefont {Kuzminov}\ and\ \citenamefont
  {Osetrova}(2000)}]{C14_Kuzminov}%
  \BibitemOpen
  \bibfield  {author} {\bibinfo {author} {\bibfnamefont {V.~V.}\ \bibnamefont
  {Kuzminov}}\ and\ \bibinfo {author} {\bibfnamefont {N.~J.}\ \bibnamefont
  {Osetrova}},\ }\href {\doibase 10.1134/1.855786} {\bibfield  {journal}
  {\bibinfo  {journal} {Physics of Atomic Nuclei}\ }\textbf {\bibinfo {volume}
  {63}},\ \bibinfo {pages} {1292} (\bibinfo {year} {2000})}\BibitemShut
  {NoStop}%
\bibitem [{\citenamefont {Wietfeldt}\ \emph {et~al.}(1995)\citenamefont
  {Wietfeldt}, \citenamefont {Norman}, \citenamefont {Chan}, \citenamefont
  {da~Cruz}, \citenamefont {Garc\'{\i}a}, \citenamefont {Haller}, \citenamefont
  {Hansen}, \citenamefont {Hindi}, \citenamefont {Larimer}, \citenamefont
  {Lesko}, \citenamefont {Luke}, \citenamefont {Stokstad}, \citenamefont
  {Sur},\ and\ \citenamefont {\ifmmode~\check{Z}\else
  \v{Z}\fi{}limen}}]{C14_Wietfeldt}%
  \BibitemOpen
  \bibfield  {author} {\bibinfo {author} {\bibfnamefont {F.~E.}\ \bibnamefont
  {Wietfeldt}}, \bibinfo {author} {\bibfnamefont {E.~B.}\ \bibnamefont
  {Norman}}, \bibinfo {author} {\bibfnamefont {Y.~D.}\ \bibnamefont {Chan}},
  \bibinfo {author} {\bibfnamefont {M.~T.~F.}\ \bibnamefont {da~Cruz}},
  \bibinfo {author} {\bibfnamefont {A.}~\bibnamefont {Garc\'{\i}a}}, \bibinfo
  {author} {\bibfnamefont {E.~E.}\ \bibnamefont {Haller}}, \bibinfo {author}
  {\bibfnamefont {W.~L.}\ \bibnamefont {Hansen}}, \bibinfo {author}
  {\bibfnamefont {M.~M.}\ \bibnamefont {Hindi}}, \bibinfo {author}
  {\bibfnamefont {R.-M.}\ \bibnamefont {Larimer}}, \bibinfo {author}
  {\bibfnamefont {K.~T.}\ \bibnamefont {Lesko}}, \bibinfo {author}
  {\bibfnamefont {P.~N.}\ \bibnamefont {Luke}}, \bibinfo {author}
  {\bibfnamefont {R.~G.}\ \bibnamefont {Stokstad}}, \bibinfo {author}
  {\bibfnamefont {B.}~\bibnamefont {Sur}}, \ and\ \bibinfo {author}
  {\bibfnamefont {I.}~\bibnamefont {\ifmmode~\check{Z}\else \v{Z}\fi{}limen}},\
  }\href {\doibase 10.1103/PhysRevC.52.1028} {\bibfield  {journal} {\bibinfo
  {journal} {Phys. Rev. C}\ }\textbf {\bibinfo {volume} {52}},\ \bibinfo
  {pages} {1028} (\bibinfo {year} {1995})}\BibitemShut {NoStop}%
\bibitem [{\citenamefont {Aprile}\ and\ \citenamefont
  {Doke}(2010{\natexlab{a}})}]{lxe_detectors}%
  \BibitemOpen
  \bibfield  {author} {\bibinfo {author} {\bibfnamefont {E.}~\bibnamefont
  {Aprile}}\ and\ \bibinfo {author} {\bibfnamefont {T.}~\bibnamefont {Doke}},\
  }\href {\doibase 10.1103/RevModPhys.82.2053} {\bibfield  {journal} {\bibinfo
  {journal} {Rev. Mod. Phys.}\ }\textbf {\bibinfo {volume} {82}},\ \bibinfo
  {pages} {2053} (\bibinfo {year} {2010}{\natexlab{a}})}\BibitemShut {NoStop}%
\bibitem [{\citenamefont {Akerib}\ \emph
  {et~al.}(2018{\natexlab{b}})\citenamefont {Akerib} \emph
  {et~al.}}]{lux_posrec}%
  \BibitemOpen
  \bibfield  {author} {\bibinfo {author} {\bibfnamefont {D.~S.}\ \bibnamefont
  {Akerib}} \emph {et~al.} (\bibinfo {collaboration} {LUX Collaboration}),\
  }\href {\doibase 10.1088/1748-0221/13/02/P02001} {\bibfield  {journal}
  {\bibinfo  {journal} {JINST}\ }\textbf {\bibinfo {volume} {13}},\ \bibinfo
  {pages} {P02001} (\bibinfo {year} {2018}{\natexlab{b}})},\ \Eprint
  {http://arxiv.org/abs/1710.02752} {arXiv:1710.02752 [physics.ins-det]}
  \BibitemShut {NoStop}%
%%CITATION = ARXIV:1710.02752;%%
\bibitem [{\citenamefont {Akerib}\ \emph
  {et~al.}(2017{\natexlab{d}})\citenamefont {Akerib} \emph
  {et~al.}}]{lux_efield}%
  \BibitemOpen
  \bibfield  {author} {\bibinfo {author} {\bibfnamefont {D.~S.}\ \bibnamefont
  {Akerib}} \emph {et~al.} (\bibinfo {collaboration} {LUX Collaboration}),\
  }\href@noop {} {\  (\bibinfo {year} {2017}{\natexlab{d}})},\ \Eprint
  {http://arxiv.org/abs/1709.00095} {arXiv:1709.00095 [physics.ins-det]}
  \BibitemShut {NoStop}%
%%CITATION = ARXIV:1709.00095;%%
\bibitem [{\citenamefont {{Kastens}}\ \emph {et~al.}(2010)\citenamefont
  {{Kastens}}, \citenamefont {{Bedikian}}, \citenamefont {{Cahn}},
  \citenamefont {{Manzur}},\ and\ \citenamefont {{McKinsey}}}]{lux_kr1}%
  \BibitemOpen
  \bibfield  {author} {\bibinfo {author} {\bibfnamefont {L.~W.}\ \bibnamefont
  {{Kastens}}}, \bibinfo {author} {\bibfnamefont {S.}~\bibnamefont
  {{Bedikian}}}, \bibinfo {author} {\bibfnamefont {S.~B.}\ \bibnamefont
  {{Cahn}}}, \bibinfo {author} {\bibfnamefont {A.}~\bibnamefont {{Manzur}}}, \
  and\ \bibinfo {author} {\bibfnamefont {D.~N.}\ \bibnamefont {{McKinsey}}},\
  }\href {\doibase 10.1088/1748-0221/5/05/P05006} {\bibfield  {journal}
  {\bibinfo  {journal} {Journal of Instrumentation}\ }\textbf {\bibinfo
  {volume} {5}},\ \bibinfo {pages} {5006} (\bibinfo {year} {2010})},\ \Eprint
  {http://arxiv.org/abs/0912.2337} {arXiv:0912.2337 [physics.ins-det]}
  \BibitemShut {NoStop}%
\bibitem [{\citenamefont {Kastens}\ \emph {et~al.}(2009)\citenamefont
  {Kastens}, \citenamefont {Cahn}, \citenamefont {Manzur},\ and\ \citenamefont
  {McKinsey}}]{lux_kr2}%
  \BibitemOpen
  \bibfield  {author} {\bibinfo {author} {\bibfnamefont {L.~W.}\ \bibnamefont
  {Kastens}}, \bibinfo {author} {\bibfnamefont {S.~B.}\ \bibnamefont {Cahn}},
  \bibinfo {author} {\bibfnamefont {A.}~\bibnamefont {Manzur}}, \ and\ \bibinfo
  {author} {\bibfnamefont {D.~N.}\ \bibnamefont {McKinsey}},\ }\href {\doibase
  10.1103/PhysRevC.80.045809} {\bibfield  {journal} {\bibinfo  {journal} {Phys.
  Rev.}\ }\textbf {\bibinfo {volume} {C80}},\ \bibinfo {pages} {045809}
  (\bibinfo {year} {2009})},\ \Eprint {http://arxiv.org/abs/0905.1766}
  {arXiv:0905.1766 [physics.ins-det]} \BibitemShut {NoStop}%
%%CITATION = ARXIV:0905.1766;%%
\bibitem [{\citenamefont {Akerib}\ \emph
  {et~al.}(2017{\natexlab{e}})\citenamefont {Akerib} \emph {et~al.}}]{lux_kr3}%
  \BibitemOpen
  \bibfield  {author} {\bibinfo {author} {\bibfnamefont {D.~S.}\ \bibnamefont
  {Akerib}} \emph {et~al.} (\bibinfo {collaboration} {LUX Collaboration}),\
  }\href {\doibase 10.1103/PhysRevD.96.112009} {\bibfield  {journal} {\bibinfo
  {journal} {Phys. Rev. D}\ }\textbf {\bibinfo {volume} {96}},\ \bibinfo
  {pages} {112009} (\bibinfo {year} {2017}{\natexlab{e}})}\BibitemShut
  {NoStop}%
\bibitem [{\citenamefont {Verbus}\ \emph {et~al.}(2017)\citenamefont {Verbus}
  \emph {et~al.}}]{lux_dd1}%
  \BibitemOpen
  \bibfield  {author} {\bibinfo {author} {\bibfnamefont {J.~R.}\ \bibnamefont
  {Verbus}} \emph {et~al.},\ }\href {\doibase 10.1016/j.nima.2017.01.053}
  {\bibfield  {journal} {\bibinfo  {journal} {Nucl. Instrum. Meth.}\ }\textbf
  {\bibinfo {volume} {A851}},\ \bibinfo {pages} {68} (\bibinfo {year}
  {2017})},\ \Eprint {http://arxiv.org/abs/1608.05309} {arXiv:1608.05309
  [physics.ins-det]} \BibitemShut {NoStop}%
%%CITATION = ARXIV:1608.05309;%%
\bibitem [{\citenamefont {Akerib}\ \emph
  {et~al.}(2016{\natexlab{c}})\citenamefont {Akerib} \emph {et~al.}}]{lux_dd2}%
  \BibitemOpen
  \bibfield  {author} {\bibinfo {author} {\bibfnamefont {D.~S.}\ \bibnamefont
  {Akerib}} \emph {et~al.} (\bibinfo {collaboration} {LUX Collaboration}),\
  }\href@noop {} {\  (\bibinfo {year} {2016}{\natexlab{c}})},\ \Eprint
  {http://arxiv.org/abs/1608.05381} {arXiv:1608.05381 [physics.ins-det]}
  \BibitemShut {NoStop}%
%%CITATION = ARXIV:1608.05381;%%
\bibitem [{\citenamefont {Boulton}\ \emph {et~al.}(2017)\citenamefont
  {Boulton}, \citenamefont {Bernard}, \citenamefont {Destefano}, \citenamefont
  {Edwards}, \citenamefont {Gai}, \citenamefont {Hertel}, \citenamefont {Horn},
  \citenamefont {Larsen}, \citenamefont {Tennyson}, \citenamefont {Wahl},\ and\
  \citenamefont {McKinsey}}]{pixey_ar37}%
  \BibitemOpen
  \bibfield  {author} {\bibinfo {author} {\bibfnamefont {E.}~\bibnamefont
  {Boulton}}, \bibinfo {author} {\bibfnamefont {E.}~\bibnamefont {Bernard}},
  \bibinfo {author} {\bibfnamefont {N.}~\bibnamefont {Destefano}}, \bibinfo
  {author} {\bibfnamefont {B.}~\bibnamefont {Edwards}}, \bibinfo {author}
  {\bibfnamefont {M.}~\bibnamefont {Gai}}, \bibinfo {author} {\bibfnamefont
  {S.}~\bibnamefont {Hertel}}, \bibinfo {author} {\bibfnamefont
  {M.}~\bibnamefont {Horn}}, \bibinfo {author} {\bibfnamefont {N.}~\bibnamefont
  {Larsen}}, \bibinfo {author} {\bibfnamefont {B.}~\bibnamefont {Tennyson}},
  \bibinfo {author} {\bibfnamefont {C.}~\bibnamefont {Wahl}}, \ and\ \bibinfo
  {author} {\bibfnamefont {D.}~\bibnamefont {McKinsey}},\ }\href
  {http://stacks.iop.org/1748-0221/12/i=08/a=P08004} {\bibfield  {journal}
  {\bibinfo  {journal} {Journal of Instrumentation}\ }\textbf {\bibinfo
  {volume} {12}},\ \bibinfo {pages} {P08004} (\bibinfo {year}
  {2017})}\BibitemShut {NoStop}%
\bibitem [{mor()}]{moravek}%
  \BibitemOpen
  \href@noop {} {}\bibinfo {note} {Moravek Biochemical Brea California 92821
  U.S.A}\BibitemShut {NoStop}%
\bibitem [{\citenamefont {Dobi}(2014)}]{attila}%
  \BibitemOpen
  \bibfield  {author} {\bibinfo {author} {\bibfnamefont {A.}~\bibnamefont
  {Dobi}},\ }\emph {\bibinfo {title} {Measurement of the Electron Recoil Band
  of the LUX Dark Matter Detector with a Tritium Calibration Source}},\ \href
  {\doibase doi:10.13016/M24P5P} {Ph.D. thesis},\ \bibinfo  {school} {Uniersity
  of Maryland} (\bibinfo {year} {2014})\BibitemShut {NoStop}%
\bibitem [{\citenamefont {Szydagis}\ \emph {et~al.}(2011)\citenamefont
  {Szydagis}, \citenamefont {Barry}, \citenamefont {Kazkaz}, \citenamefont
  {Mock}, \citenamefont {Stolp}, \citenamefont {Sweany}, \citenamefont
  {Tripathi}, \citenamefont {Uvarov}, \citenamefont {Walsh},\ and\
  \citenamefont {Woods}}]{nest1}%
  \BibitemOpen
  \bibfield  {author} {\bibinfo {author} {\bibfnamefont {M.}~\bibnamefont
  {Szydagis}}, \bibinfo {author} {\bibfnamefont {N.}~\bibnamefont {Barry}},
  \bibinfo {author} {\bibfnamefont {K.}~\bibnamefont {Kazkaz}}, \bibinfo
  {author} {\bibfnamefont {J.}~\bibnamefont {Mock}}, \bibinfo {author}
  {\bibfnamefont {D.}~\bibnamefont {Stolp}}, \bibinfo {author} {\bibfnamefont
  {M.}~\bibnamefont {Sweany}}, \bibinfo {author} {\bibfnamefont
  {M.}~\bibnamefont {Tripathi}}, \bibinfo {author} {\bibfnamefont
  {S.}~\bibnamefont {Uvarov}}, \bibinfo {author} {\bibfnamefont
  {N.}~\bibnamefont {Walsh}}, \ and\ \bibinfo {author} {\bibfnamefont
  {M.}~\bibnamefont {Woods}},\ }\href@noop {} {\bibfield  {journal} {\bibinfo
  {journal} {Journal of Instrumentation}\ }\textbf {\bibinfo {volume} {6}},\
  \bibinfo {pages} {P10002} (\bibinfo {year} {2011})}\BibitemShut {NoStop}%
\bibitem [{\citenamefont {Lenardo}\ \emph {et~al.}(2015)\citenamefont
  {Lenardo}, \citenamefont {Kazkaz}, \citenamefont {Manalaysay}, \citenamefont
  {Mock}, \citenamefont {Szydagis},\ and\ \citenamefont {Tripathi}}]{nest3}%
  \BibitemOpen
  \bibfield  {author} {\bibinfo {author} {\bibfnamefont {B.}~\bibnamefont
  {Lenardo}}, \bibinfo {author} {\bibfnamefont {K.}~\bibnamefont {Kazkaz}},
  \bibinfo {author} {\bibfnamefont {A.}~\bibnamefont {Manalaysay}}, \bibinfo
  {author} {\bibfnamefont {J.}~\bibnamefont {Mock}}, \bibinfo {author}
  {\bibfnamefont {M.}~\bibnamefont {Szydagis}}, \ and\ \bibinfo {author}
  {\bibfnamefont {M.}~\bibnamefont {Tripathi}},\ }\href {\doibase
  10.1109/TNS.2015.2481322} {\bibfield  {journal} {\bibinfo  {journal} {IEEE
  Trans. Nucl. Sci.}\ }\textbf {\bibinfo {volume} {62}},\ \bibinfo {pages}
  {3387} (\bibinfo {year} {2015})},\ \Eprint {http://arxiv.org/abs/1412.4417}
  {arXiv:1412.4417 [astro-ph.IM]} \BibitemShut {NoStop}%
%%CITATION = ARXIV:1412.4417;%%
\bibitem [{\citenamefont {Aprile}\ and\ \citenamefont
  {Doke}(2010{\natexlab{b}})}]{aprile_doke_LXe}%
  \BibitemOpen
  \bibfield  {author} {\bibinfo {author} {\bibfnamefont {E.}~\bibnamefont
  {Aprile}}\ and\ \bibinfo {author} {\bibfnamefont {T.}~\bibnamefont {Doke}},\
  }\href {\doibase 10.1103/RevModPhys.82.2053} {\bibfield  {journal} {\bibinfo
  {journal} {Rev. Mod. Phys.}\ }\textbf {\bibinfo {volume} {82}},\ \bibinfo
  {pages} {2053} (\bibinfo {year} {2010}{\natexlab{b}})}\BibitemShut {NoStop}%
\bibitem [{\citenamefont {Platzman}(1961)}]{platzman}%
  \BibitemOpen
  \bibfield  {author} {\bibinfo {author} {\bibfnamefont {R.}~\bibnamefont
  {Platzman}},\ }\href@noop {} {\bibfield  {journal} {\bibinfo  {journal} {The
  International Journal of Applied Radiation and Isotopes}\ }\textbf {\bibinfo
  {volume} {10}},\ \bibinfo {pages} {116} (\bibinfo {year} {1961})}\BibitemShut
  {NoStop}%
\bibitem [{\citenamefont {Dahl}(2009)}]{dahl}%
  \BibitemOpen
  \bibfield  {author} {\bibinfo {author} {\bibfnamefont {C.~E.}\ \bibnamefont
  {Dahl}},\ }\emph {\bibinfo {title} {The physics of background discrimination
  in liquid xenon, and first results from Xenon10 in the hunt for WIMP dark
  matter.}},\ \href@noop {} {Ph.D. thesis},\ \bibinfo  {school} {Princeton
  University} (\bibinfo {year} {2009})\BibitemShut {NoStop}%
\bibitem [{\citenamefont {Doke}\ \emph {et~al.}(2002)\citenamefont {Doke},
  \citenamefont {Hitachi}, \citenamefont {Kikuchi}, \citenamefont {Masuda},
  \citenamefont {Okada},\ and\ \citenamefont {Shibamura}}]{doke2002}%
  \BibitemOpen
  \bibfield  {author} {\bibinfo {author} {\bibfnamefont {T.}~\bibnamefont
  {Doke}}, \bibinfo {author} {\bibfnamefont {A.}~\bibnamefont {Hitachi}},
  \bibinfo {author} {\bibfnamefont {J.}~\bibnamefont {Kikuchi}}, \bibinfo
  {author} {\bibfnamefont {K.}~\bibnamefont {Masuda}}, \bibinfo {author}
  {\bibfnamefont {H.}~\bibnamefont {Okada}}, \ and\ \bibinfo {author}
  {\bibfnamefont {E.}~\bibnamefont {Shibamura}},\ }\href
  {http://stacks.iop.org/1347-4065/41/i=3R/a=1538} {\bibfield  {journal}
  {\bibinfo  {journal} {Japanese Journal of Applied Physics}\ }\textbf
  {\bibinfo {volume} {41}},\ \bibinfo {pages} {1538} (\bibinfo {year}
  {2002})}\BibitemShut {NoStop}%
\end{thebibliography}


\end{document}